\documentclass[fleqn,usenatbib]{mnras}

\usepackage{newtxtext,newtxmath}

\usepackage[T1]{fontenc}
\usepackage{ae,aecompl}

\usepackage{graphicx}	
\usepackage{amsmath}	
\usepackage{amssymb}	
\usepackage{gensymb}

\title[Asteroid impacts on terrestrial planets]{Asteroid impacts on terrestrial planets: The effects of super--Earths and the role of the $\nu_6$ resonance}

\author[Smallwood et al.]{
Jeremy L. Smallwood,\thanks{E-mail: smallj2@unlv.nevada.edu}
Rebecca G. Martin,
Stephen Lepp
and Mario Livio
\\
Department of Physics and Astronomy, University of Nevada, Las Vegas, 4505 South Maryland Parkway, Las Vegas, NV 89154, USA
}
\date{Accepted XXX. Received YYY; in original form ZZZ}

\pubyear{2017}

\begin{document}
\label{firstpage}
\pagerange{\pageref{firstpage}--\pageref{lastpage}}
\maketitle

\begin{abstract}
With N--body simulations of a planetary system with an asteroid belt we investigate how the asteroid impact rate on the Earth is affected by the architecture of the planetary system. We find that the $\nu_6$ secular resonance plays an important role in the asteroid collision rate with the Earth.  Compared to exoplanetary systems, the solar system is somewhat special in its lack of a super--Earth mass planet in the inner solar system. We therefore first consider the effects of the presence of a super--Earth in the terrestrial planet region.  We find a significant effect for super--Earths with a mass of around $10\,\rm M_\oplus$ and a separation greater than about $0.7\,\rm AU$.  For a super-Earth that is interior to the Earth's orbit, the number of asteroids colliding with Earth increases the closer the super--Earth is to the Earth's orbit. This is the result of multiple secular resonance locations causing more asteroids to be perturbed onto Earth-crossing orbits. When the super-Earth is placed exterior to Earth's orbit, the collision rate decreases substantially because the $\nu_6$ resonance no longer exists in the asteroid belt region. We also find that changing the semi--major axis of Saturn leads to a significant decrease in the asteroid collision rate, while increasing its mass increases the collision rate. These results may have implications for the habitability of exoplanetary systems. 
\end{abstract}




\section{Introduction}
Asteroidal impacts on terrestrial planets may be important for planet habitability. Water must have been delivered to the surface of the Earth, potentially  through asteroid impacts \citep{MorbidelliChambers2000,RaymondQuinn2007}, even though comet collisions and the interaction between the magma and atmosphere have also been suggested as sources of the planet's water \citep{Lunine2006,GendaIkoma2008}. Large moons may be produced through asteroid collisions \citep{CanupAsphaug2001, Canup2012}. Our Moon plays an intricate role in the Earth's habitability by stabilizing the rotation axis of the Earth, preventing weather extremes due to chaotic motion.  According to some hypotheses, life itself may have been delivered to Earth by asteroids \citep{Cronin1983, castillo2008, Houtkooper2011}. Important heavy elements needed for the existence of life may have also been delivered by asteroid impacts \citep{Willbold2011}. The \textit{rate} of asteroid impacts also can have a significant effect on a planet's habitability. A high rate of impacts could have led to a highly cratered planet, not hospitable to life. On the other hand, too few impacts could suppress the delivery of essential elements. The evolution of life on Earth shows that the presence of an asteroid belt could be necessary (e.g. through mass extinctions) even for the appearance of complex life and intelligence in an exosolar planetary system.

Asteroid belts, if they form at all, form most likely interior to giant planets \citep{Morales2011, Martin2013}. This coincides with the location of the water snow line, the radius outside of which water is found in a solid form \citep{Lecar2006}. Giant planets form outside of the snow line and the terrestrial planets form inward of it \citep{Raymond2009}. While the gas disk is present, the eccentricities and inclinations of asteroids are damped by tidal interactions with the protoplanetary gas disk \citep{Ward1989,Ward1993,Artymowicz1993,AgnorWard2002,Kominami2002}. The lifetime of the gas disk is typically a few million years \citep{Haisch2001}, after which photoevaporation disperses the disk. A pressure gradient arises when stellar radiation heats the disk, allowing the thermal energy of the gas to exceed its gravitational binding energy \citep{Alexander2006,Armitage2013}. Once the gas disk is removed, gravitational perturbations cause a clearing of asteroids at many resonances locations \citep{Morbidelli1995,Gladman1997,Morbidelli1998,PetitMorbidelli2001, Bro2008, Minton2010, Chrenko2015} leading to the formation of potential Earth impactors \citep{Morbidelli1999,Strom2005}. 

\begin{figure} 
\includegraphics[width=8.4cm]{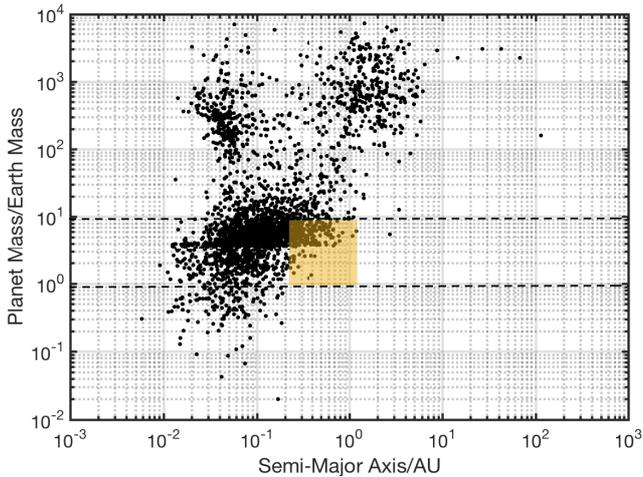}
\caption{Planet mass and semi-major axis of observed exoplanets.  The area between the two black-dotted lines contains the range of super-Earth masses ($1\,\rm M_\oplus < \rm M_p < 10\,\rm M_\oplus$) used in simulations described in Section 2. The transparent yellow box denotes the observed super-Earths with a semi-major axis in the inner solar system. The data are from exoplanets.org \citep{Han2014}. }
\label{exodata}
\end{figure}

Since Jupiter and Saturn are the largest planets in the solar system, they are the main driving force for the dynamic evolution of the asteroid belt. The effect that Jupiter had on the asteroid collision rate on the Earth is debated \citep{Horner2008,Horner2012}. The asteroid belt we observe today is roughly $0.1 \%$ in mass of the original asteroid belt formed in the beginning of the solar system. In a mean-motion resonance, the ratio of the orbital periods of two objects is an integer ratio \cite[e.g.][]{Armitage2013}. Several of the  well known mean-motion resonances in our solar system are the Kirkwood gaps \citep{Dermott1983,Moons1996,Vrbik1996,Obrien2007}, which are found within the asteroid belt. Secular resonances arise when the apsidal or nodal precession rates of two objects orbiting a central object are close \citep{Yoshikawa1987}. For a review of the dynamics of secular resonances and mean-motion resonances, see \cite{Froeschle1986} and \cite{Moons1996}, respectively.

The most prevalent secular resonance in our solar system in the $\nu_6$ resonance \citep{Ito2006, Minton2011}, which occurs between the apsidal precession of the asteroids and Saturn.  The outer edge of the $\nu_{6}$ resonance location sets the inner boundary for our asteroid belt, which is approximately located at $2\,\rm AU$. A large fraction of near Earth asteroids originates from the inner main belt rather than the middle or outer belt \citep{bottke2002}. Many of these asteroids within the inner main belt are injected in the $\nu_6$ resonance and end up reaching Earth-crossing orbits \citep{Morbidelli1994,Bottke2000}.

When asteroids encounter a resonance, their eccentricities are increased to a point where they are either ejected from the system or collide with a planet or the central star. These perturbations begin about $10^6$ years after the disk is dispersed \citep{Morbidelli1994,ItoTanikawa1999}. Each resonance has a width, in semi-major axis, over which it operates \citep{Dermott1983}. Asteroids that fall within the resonance width will undergo perturbations causing their eccentricities to increase. Regions in which the resonance widths overlap are known as chaotic regions \citep{Murray1997, Murray1999}, and there almost all of the asteroids will be cleared out.  The region where Jupiter's resonances overlap is what determines the outer edge of the asteroid belt at about $3.3\,\rm AU$.

The importance of the $\nu_6$ resonance is apparent through observations of craters in the solar system. The Late Heavy Bombardment (LHB) event occurred early in the evolution of the solar system. During this period, many lunar craters were formed from a large barrage of comets and asteroids. The LHB is assumed to have ended about $3.7$ to $3.8$ billion years ago. \cite{Bottke2012} found that the majority of these asteroids originated from the hypothetical extension of the asteroid belt, the E-belt, which was located between $1.7$ and $2.1\, \rm AU$. This region underwent instabilities due to giant planet migration and is now depleted except for a family of high inclination asteroids.

Observations of exoplanetary systems show an abundance of systems that possess a super-Earth  \citep{Borucki2010,Borucki2011,Batalha2013}, a planet with a mass in the range of $1-10M_{\oplus}$ \citep{Valencia2007}. Fig.~\ref{exodata} shows the planet mass and semi--major axis of observed exoplanets. The area between the two black--dotted lines contains the observed super--Earths. The transparent yellow box denotes the observed super--Earths with a semi-major axis located at orbital radii in the range of the terrestrial planets in the solar system. The vast majority of observed super--Earths are close to their parent star, although this is most likely a selection effect \citep{Chiang2013}. In a recent work, \cite{Martin2015} have shown that perhaps the main characteristic that distinguishes the solar system from the observed exoplanetary systems is the absence of super--Earths. Consequently, it is very interesting to examine the effects that the presence of a super--Earth would have had on asteroid impacts on Earth. Since the evolution of a planetary system is a chaotic process \citep{Malhotra1999}, adding a super-Earth can greatly influence the architecture of an exoplanetary system. There are likely additional mechanisms that operate in exoplanetary systems that result in the migration of planets, through the gas disk \citep{Armitage2013}, through the planetesimal disk \citep{Wyatt2003,Gomes2004}, via secular processes such as the Kozai--Lidov mechanism \citep{Kozai1962,Lidov1962,Wu2003,Takeda2005,Martin2016} and through planet--planet scattering \citep{Ford2008,Dawson2013}. The Kepler data suggest that there is a overabundance of planetary systems with only one planet \citep{Lissauer2011,Hansen2013} and this may be due to the destructive motion of the planet \citep[e.g.][]{Morton2014}.

In this work we are interested in multi--planet exoplanetary  systems in which there has not been violent processes that might have destroyed an asteroid belt and prevented the formation of terrestrial planets early in the lifetime of the system. Specifically, we investigate how the architecture of the solar system affects asteroid impacts on the Earth. In Section~2 we describe the N--body simulations that we use to model the evolution of an asteroid belt in a planetary system. In Section~3 we consider the effect of changing the structure of the inner solar system. In particular, we examine the effect the presence of a super--Earth would have had. We find that the $\nu_6$ resonance plays an important role in causing asteroid collisions with the Earth. In Section~4 we consider how the giant planets in the outer solar system affect the location and strength of the $\nu_6$ resonance. We draw our conclusions in Section~5.

\section{N--body simulations}

We use the hybrid symplectic integrator in  the orbital dynamics package, {\sc mercury}, to model the structure of the asteroid belt and the asteroid impact rate on the Earth in the presence of a super--Earth. {\sc mercury} uses N-body integrations to calculate the orbital evolution of objects moving in the gravitational field of a large body \citep{Chambers1999}. We simulate the motion of a super-Earth, Earth, Jupiter, Saturn, and a distribution of asteroids orbiting a central object. The asteroids interact gravitationally with the planets and the star but do not interact with one another. We calculate the  evolution of each asteroid orbit for a duration of ten million years. 

The total mass observed in  the solar system's asteroid belt is
about $5 \times 10^{-4} M_{\oplus}$, with about $80\%$ of the mass contained in the three largest asteroids (Ceres, Pallas, and Vesta). Today, there are over $10,000$ asteroids with high accuracy measurements of the semi-major axis, with the mean being $<a> = 2.74 \pm 0.616$ AU. The mean eccentricity is $<e> = 0.148 \pm 0.086$ and the mean inclination is $<i> = 8.58 \degree \pm 6.62\degree$ \citep{MurrayBook2000}. However, the exact initial structure of the distribution of asteroids within an asteroid belt immediately after the dispersal of the protoplanetary disk is not fully understood. We assume that the asteroid distribution is sampled from a uniform distribution \citep{Lecar1997}. The semi-major axis of each asteroid is given by

\begin{equation}
a_{\rm i} = (a_{\rm max} - a_{\rm min}) \times \chi_r + a_{\rm min},
\end{equation}
where $a_{\rm min} = 1.558\,\rm AU$ is the inner boundary of the distribution, $a_{\rm max} = 4.138\,$AU is the outer boundary, and $\chi_r$ is a randomly generated number between $0$ and $1$. The inner and outer boundaries are determined using the structure of our own solar system, $a_{\rm min}$ is three Hill radii beyond the semi-major axis of Mars and $a_{\rm max}$ is located at three Hill radii interior to Jupiter's orbit. The Hill radius is given by
\begin{equation}
R_{\rm H} = a_{\rm p}\bigg(\frac{M_{\rm planet}}{3M_{\rm star}}\bigg)^{\frac{1}{3}}
\end{equation}
where $a_{\rm p}$ is the semi-major axis of the planet and $M_{\rm planet}$ is the mass of the planet.
Generally, three Hill radii is the planet's gravitational reach and thus no asteroids would likely be located within this region \citep{Gladman1993,Chatterjee2008,Morrison2015}.

The orbit of each asteroid is defined by six orbital elements, $a$, $i$, $e$, $n$, $g$ and $M_{\rm a}$.  The semi-major axis, $a$, is distributed uniformly in the range $a_{\rm min}<a<a_{\rm max}$, the inclination, $i$, is randomly allocated from the range $0-10\degree$, the eccentricity, $e$, is randomly generated from the range $0.0 - 0.1$. The longitude of the ascending node, $n$, the argument of perhelion, $g$, and the mean anomaly, $M_{a}$, are all uniformly randomly sampled from the range $0-360\degree$. The longitude of the ascending node, $n$, is defined as the angle from a reference direction to the ascending node, $g$ is the angle from the ascending node to the object's periastron, and $M_{\rm a}$ is the angular distance from the periastron. The present-day orbital elements for each of the planets are set as the initial parameters for the planets since the solar system is stable over long timescales \citep{Duncan1998,Ito2002}.

The asteroids in our simulations are point particles that do not interact gravitationally with one another. We may neglect this interaction because the timescale for asteroid-asteroid collisional interaction is much longer than the timescale for the action of perturbations by resonance effects. The timescale for resonant effects is on the order of $\sim 1 \, \rm Myr$, whereas some of the largest asteroids have collisional timescales that are of the order of magnitude of the age of the solar system \citep{Dohnanyi1969}.

First, we checked the scalability of our results with the number of asteroids initially in the belt. We physically inflated the radius of the Earth to be $2\times 10^6\,\rm km$, in order to enhance the number of asteroids on Earth-crossing orbits. We ran a simulation of Earth, Jupiter, Saturn, and a uniformly distributed asteroid belt.  The simulation ran for ten million years, with a timestep of eight days with an accuracy parameter of $1 \times 10^{-12}$. The accuracy parameter measures approximately how much error per step the variable-timestep symplectic algorithm will tolerate. We ran three simulations with $10^3$, $10^4$ and $10^5$ asteroids.  We found that the number of asteroid collisions scaled linearly with the initial number of asteroids, as we would expect. Thus, to ensure faster simulation times and significant results we took the total number of asteroids in the rest of our work to be $10^4$. 

\begin{figure} 
\includegraphics[width=8.4cm]{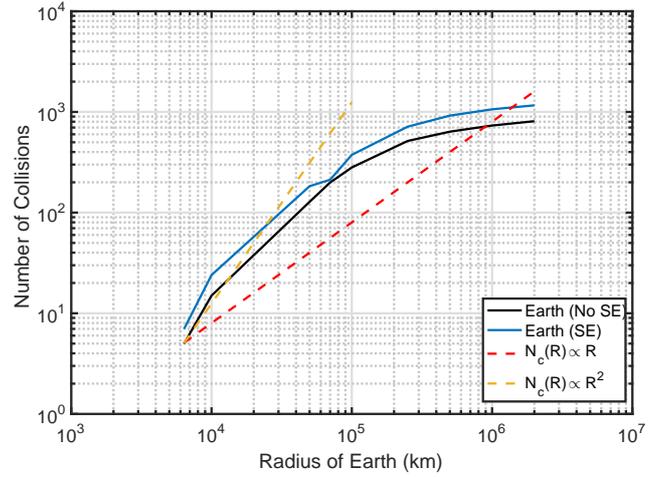}
\caption{A log--log plot showing how the number of asteroid collisions scales with the inflated radius of the Earth, $R$. The black line shows how the total number of Earth collisions by asteroids ($N_c$) scales with changing radius of our simulated inflated Earth. The blue lines shows the number of Earth collisions in the presence of a $10\,M_\oplus$ super--Earth at semi--major axis $0.8\,AU$ as a function of inflated Earth radii. The red-dotted line represents the line  $N_c(R) \propto R$ and the yellow-dotted line represents the line  $N_c(R) \propto R^2$. }
\label{scale}
\end{figure}

\begin{table*}
	\centering
	\caption{Number of asteroidal outcomes for various simulated solar system architectures involving a super--Earth (SE), Earth, Jupiter (J), and Saturn (S). Note that since the size of the Earth has been inflated, the number of Earth impacts cannot be compared to the other outcomes. We only make comparison between the number of Earth impacts between different simulations.}
	\label{tab:example_table}
	\begin{tabular}{cccccccc} 
		\hline
		Simulation Name & SE Mass & SE Semi-major Axis & Earth Impacts &J/S/SE Impacts & Star Collision & Ejected & Remaining\\
        & $\rm M_\oplus$ & AU & & & & \\
		\hline
run1 & --- & --- & 808 & 55 & 2 & 1112 & 8023 \\
\hline
run2 & $5$ & $0.22$ & 803 & 42 & 1 & 1135 & 8019 \\
run3 & $5$ & $0.40$ & 805 & 49 & 1 & 1121 & 8024 \\
run4 & $5$ & $0.50$ & 826 & 45 & 3 & 1138 & 7988 \\
run5 & $5$ & $0.60$ & 857 & 40 & 4 & 1115 & 7984 \\
run6 & $5$ & $0.70$ & 877 & 47 & 1 & 1131 & 7944 \\
run7 & $5$ & $0.80$ & 942 & 41 & 1 & 1121 & 7895 \\
run8 & $5$ & $1.20$ & 394 & 51 & 1 & 1169 & 8385 \\
run9 & $5$ & $1.40$ & 569 & 71 & 2 & 1221 & 8138 \\
\hline
run10 & $10$ & $0.22$ & 798 & 63 & 1 & 1125 & 8013 \\
run11 & $10$ & $0.40$ & 816 & 51 & 1 & 1122 & 8010 \\
run12 & $10$ & $0.50$ & 874 & 53 & 3 & 1106 & 7964 \\
run13 & $10$ & $0.60$ & 935 & 40 & 2 & 1118 & 7905 \\
run14 & $10$ & $0.70$ & 978 & 32 & 2 & 1143 & 7845 \\
run15 & $10$ & $0.80$ & 1236 & 43 & 3 & 1126 & 7592 \\
run16 & $10$ & $1.20$ & 310 & 62 & 1 & 1166 & 8461 \\
run17 & $10$ & $1.40$ & 615 & 77 & 1 & 1239 & 8068  \\
\hline
run18 & $1$ & $0.80$ & 801 & 33 & 1 & 1157 & 8008 \\
run19 & $2$ & $0.80$ & 843 & 58 & 2 & 1125 & 7972 \\
run20 & $3$ & $0.80$ & 868 & 46 & 0 & 1147 & 7939 \\
run21 & $4$ & $0.80$ & 913 & 52 & 5 & 1116 & 7914 \\
		\hline
	\end{tabular}
    \label{table1}
\end{table*}

Next, in Fig.~\ref{scale}, we examined how the number of asteroid collisions with the Earth scales with the radius of the inflated Earth.  We ran two series of simulations that included Jupiter, Saturn, the asteroid belt, and the inflated Earth. The first set does not include a Super--Earth while the second set incorporated a super--Earth of mass $10\,M_\odot$ at orbital separation $0.8\,AU$.  In each series of these simulations we change the radius of the inflated Earth and measure the resulting asteroid collisions with the Earth.  If the Earth has a radius of $1\,R_\oplus$ then the number of collisions throughout the simulation is only a few.  The simulation with the super--Earth always produces a higher rate of collisions than the simulation without a super--Earth and so the effects of the super--Earth occur at all inflated-Earth radii. Thus, inflating the Earth allows for the detection of trends within the asteroid belt. In order to get statistically significant results we inflated the size of the Earth to $2 \times 10^6\, \rm km$. Note that since the size of the Earth is inflated, we cannot compare the absolute numbers of asteroid collisions with the Earth with any other outcomes such as collisions with other planets or ejections. We are only able to compare the relative numbers of collisions with an object between simulations.

\begin{figure*} \centering
\includegraphics[width=8.7cm]{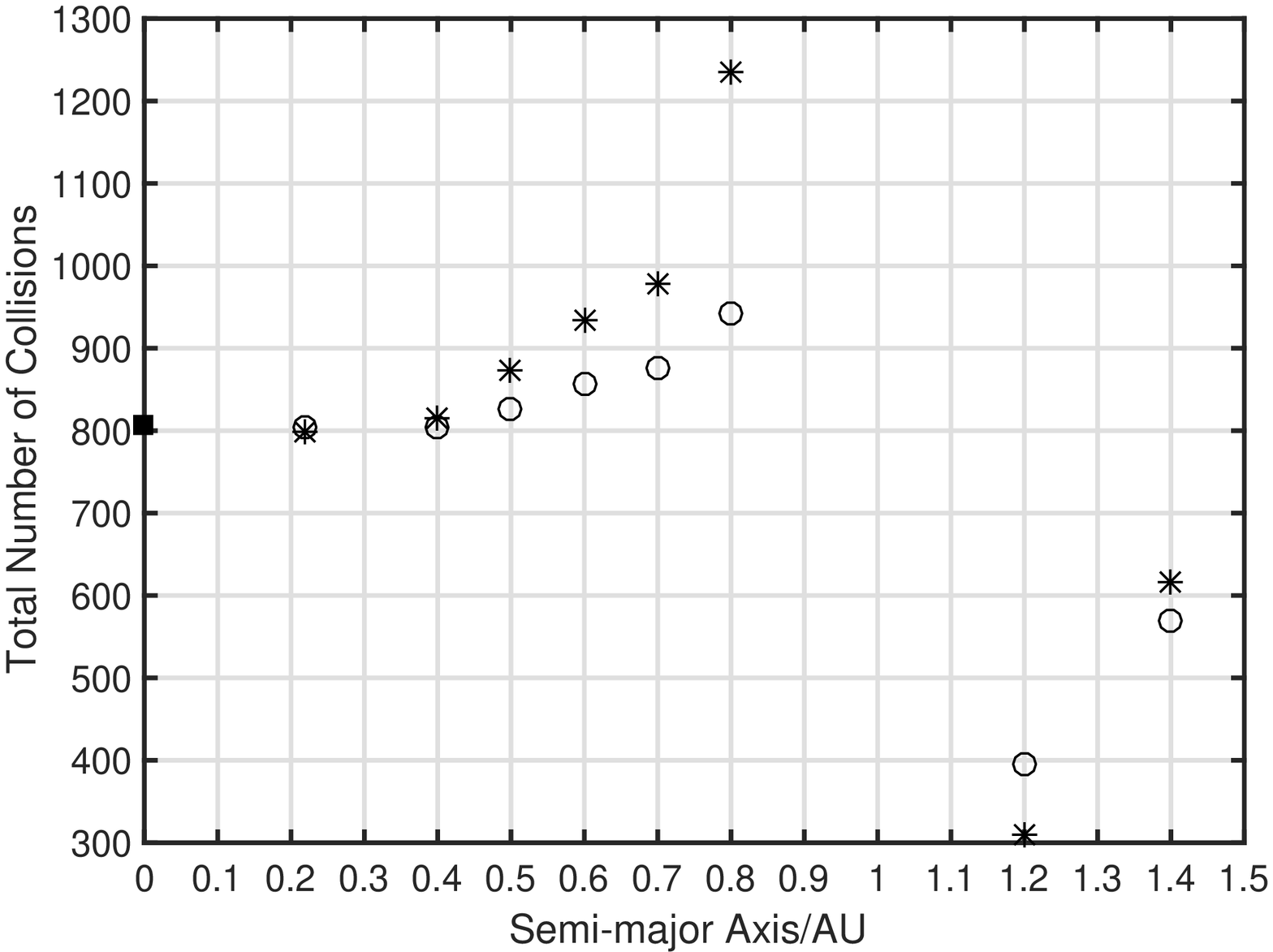}
\includegraphics[width=8.7cm]{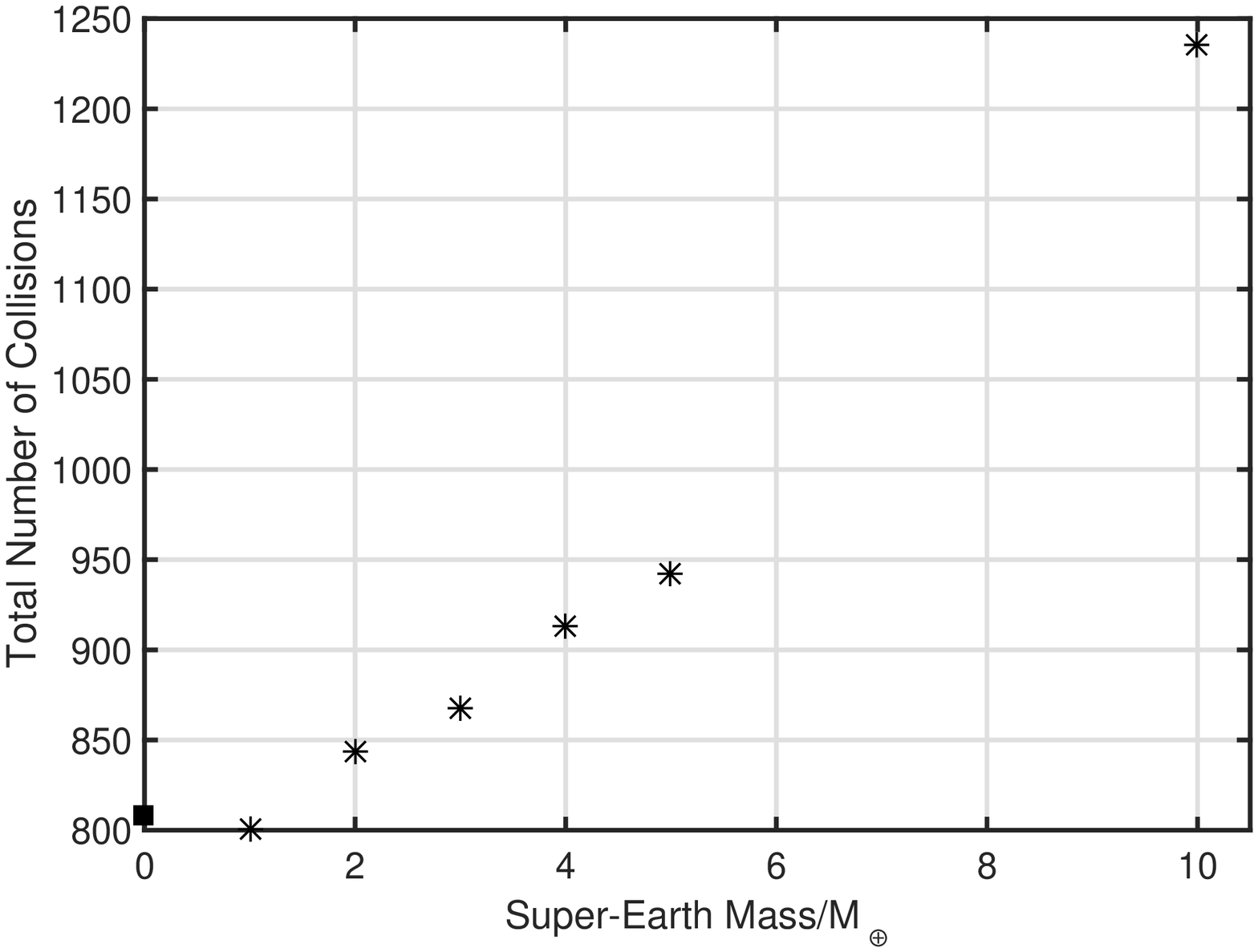}
\caption{\textit{Left Panel:} Total number of asteroid collisions with the Earth as a function of the semi--major axis of the super--Earth. The simulations with a $5M_{\oplus}$ super-Earth are denoted by hollow circles and the simulations with a $10M_{\oplus}$ super-Earth are represented by the star symbols. The square represents the simulation of the standard solar system (without a super-Earth). \textit{Right Panel:} Total number of asteroid collisions with the Earth as a function of super--Earth mass in units of Earth masses ($\rm M_{\oplus}$) at semi-major axis $a_{\rm SE} = 0.8\, \rm AU$.}
\label{number}
\end{figure*}

\section{The architecture of the inner solar system}

We first investigate the influence of the architecture of the inner solar system on the asteroid collisions on Earth. In particular, since \cite{Martin2015} identified the absence of super--Earths as perhaps the most important architectural element that distinguishes the solar system from other exoplanetary systems, we vary the mass and semi--major axis of an additional super--Earth in the inner solar system.  The super--Earth's inclination and eccentricity are initially set to zero. 

\begin{figure*} \centering
\includegraphics[width=8.7cm]{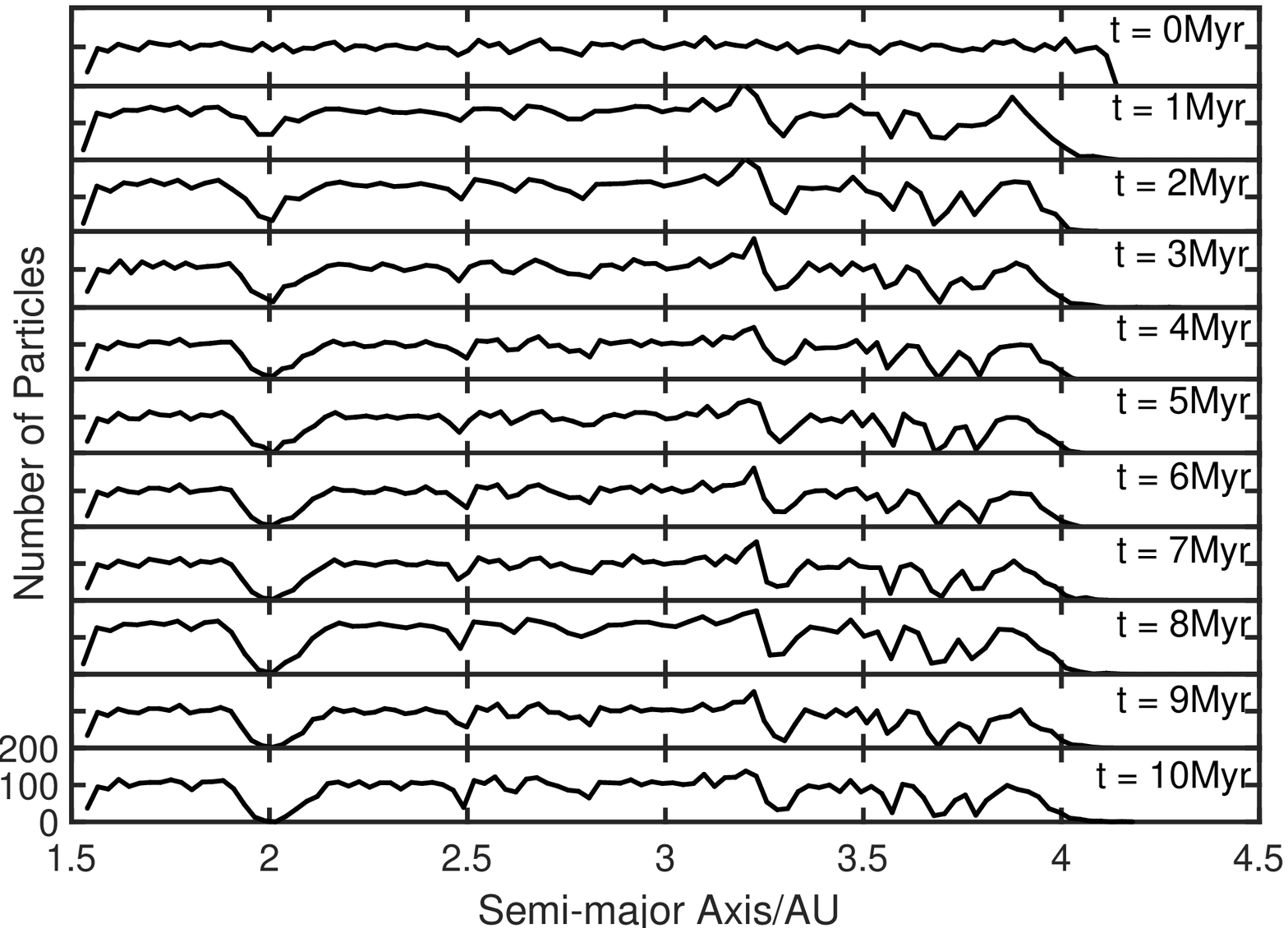}
\includegraphics[width=8.7cm]{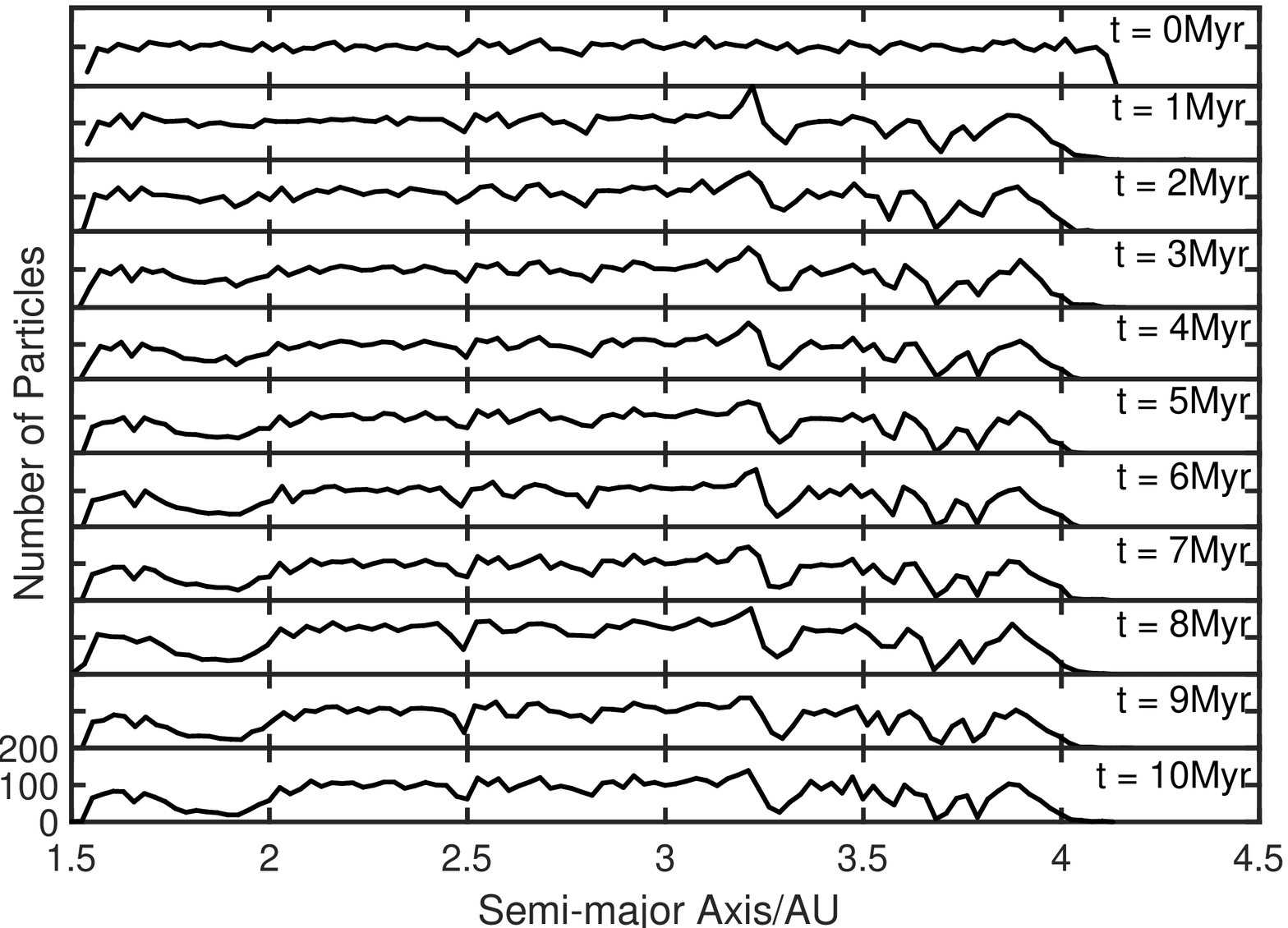}
\includegraphics[width=8.7cm]{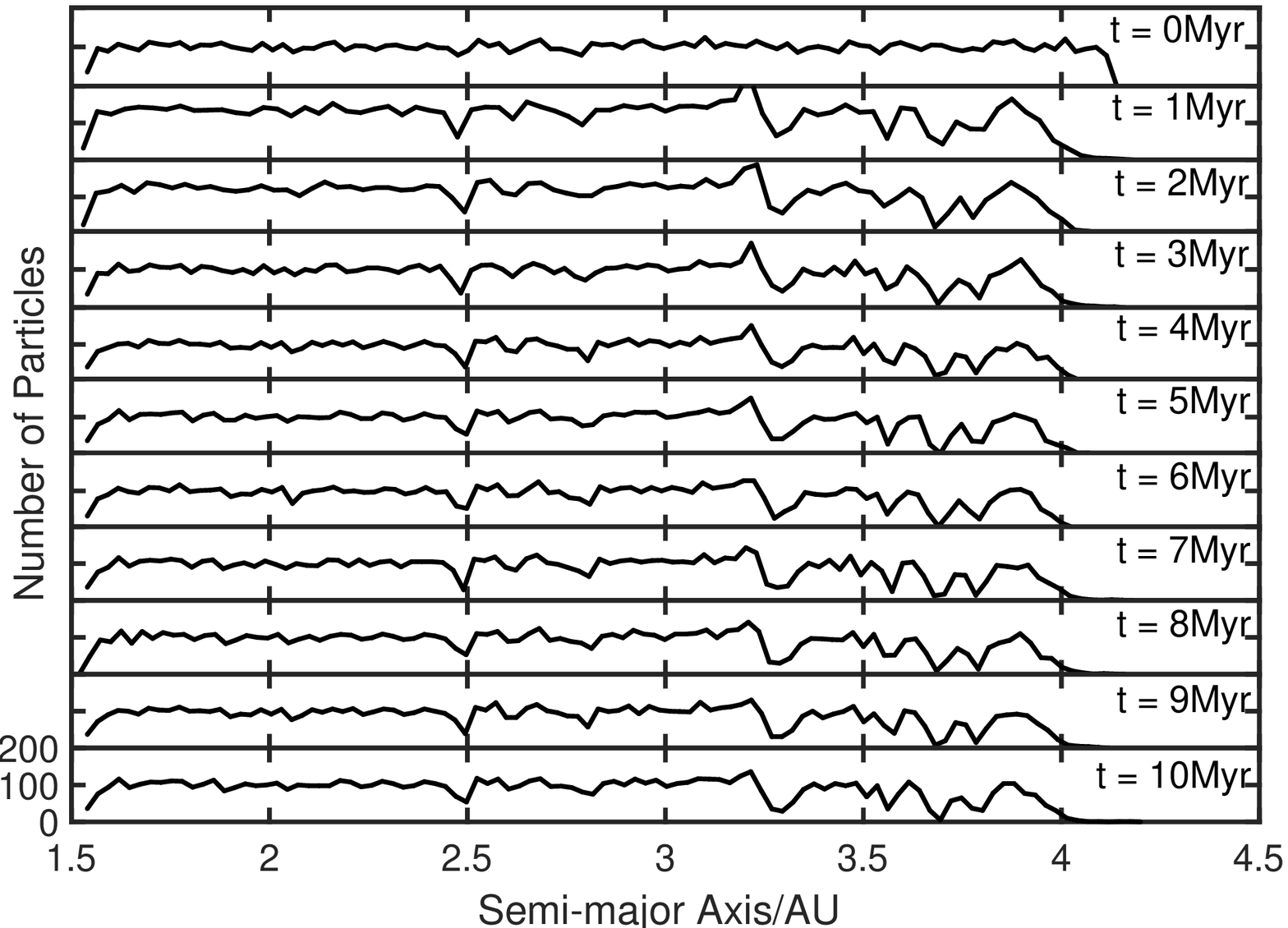}
\includegraphics[width=8.7cm]{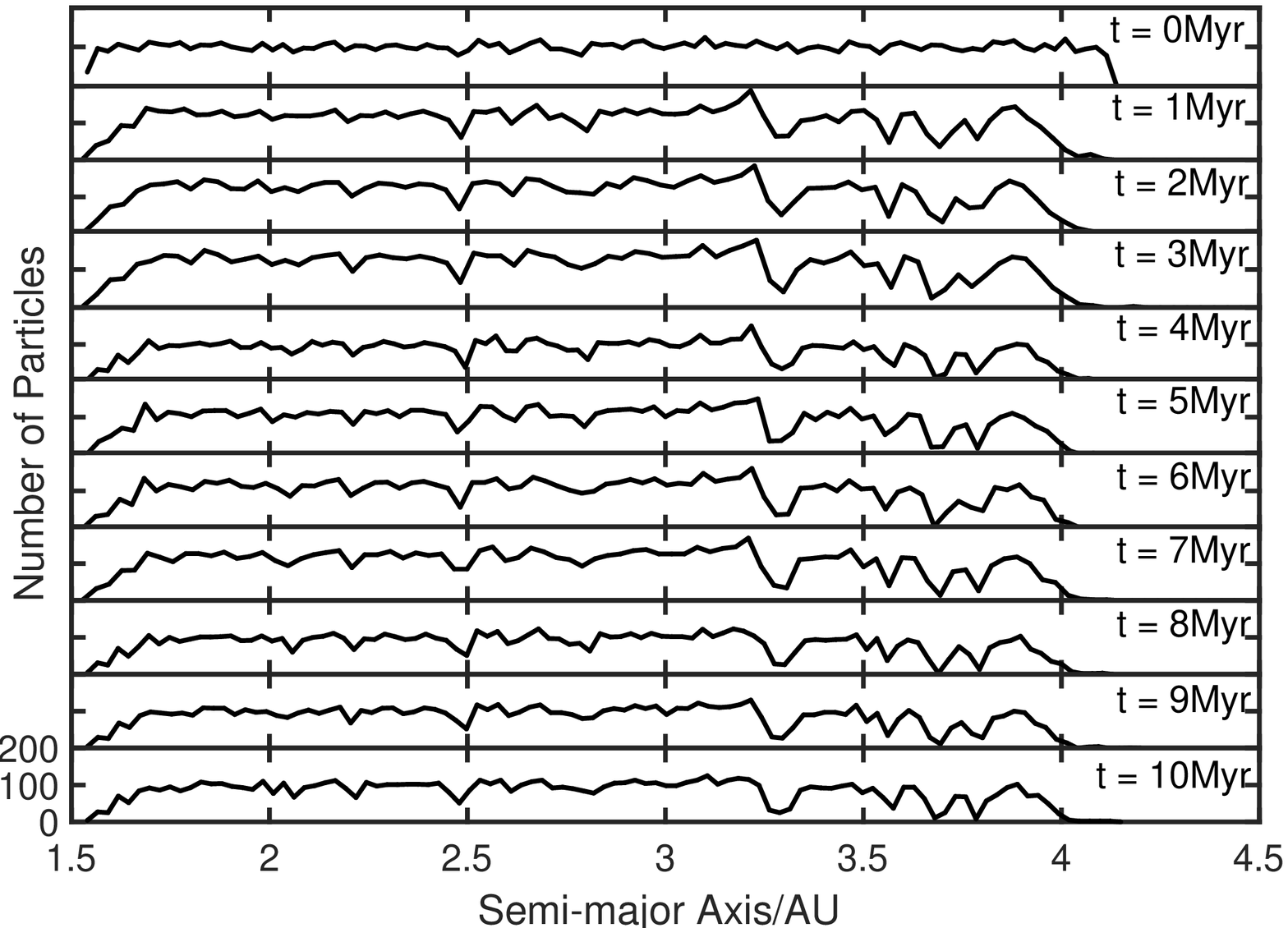}
\caption{Evolution of the asteroid distribution over a span of 10 million years. The asteroid distribution is located in a system containing Earth, Jupiter, Saturn, and a super--Earth (except for the control simulation at top left). The asteroids are initially sampled from a uniform distribution in semi-major axis values, represented at $t = 0 \, \rm Myr$. \textit{Top-Left Panel:} Evolution of asteroid distribution without a super-Earth. \textit{Top-Right Panel:} Evolution of asteroid distribution with a $10M_{\oplus}$ super-Earth located at a semi-major axis of $0.8 \, \rm AU$. \textit{Bottom-Left Panel:} Evolution of asteroid distribution with a $10M_{\oplus}$ super-Earth located at $1.2 \, \rm AU$. \textit{Bottom-Right Panel:} Evolution of asteroid distribution with a $10M_{\oplus}$ super-Earth located at $1.4 \, \rm AU$.}
\label{belt}
\end{figure*}

\subsection{Total number of Earth collisions}

In table~\ref{table1} we show the parameters of twenty-one simulations and the summary of the asteroid collisional outcomes at a time of $10\,\rm Myr$.  We used super--Earth masses in the range $1-10\,\rm M_\oplus$ and semi--major axis in the range $0.2-1.4\,\rm AU$. In each simulation we determined the total number of asteroid impacts on the Earth, impacts on the other planets, impacts on the star, the number of asteroids ejected from the system, and the number of asteroids remaining within the distribution. An asteroid is considered to have been ejected if its semi--major axis exceeds $100\,\rm AU$. Run1  from table~\ref{table1} represents our current solar system as it includes Earth, Jupiter, and Saturn but does not include a super--Earth planet. All of the simulations possessing a super--Earth were compared with that for our current solar system. Since we have inflated the size of the Earth in order to enhance the number of collisions, we compare only collision rates with the Earth between simulations. We cannot compare the number of collisions with the Earth to the other outcomes of collisions with other bodies or ejections.  We include the other outcomes only for completeness. 

Fig.~\ref{number} shows the total number of asteroid collisions with the Earth (left panel) and total number of collisions with the Earth as a function of super--Earth mass with an orbital separation of $0.8\, \rm AU$ (right panel) after a time of $10\,\rm Myr$. A super--Earth that is located interior to Earth's orbit increases the asteroid impact rate on the Earth compared to a system without a super--Earth. A super-Earth located exterior to Earth's orbit, decreases the asteroid impact rate compared to a system without a super--Earth. A $\rm 10 \, M_{\oplus}$ super-Earth located at a semi-major axis of $0.8\, \rm AU$ causes the largest number of impacts on Earth, whereas a $\rm 10 \, M_{\oplus}$ super-Earth located at a semi-major axis of $1.20\, \rm AU$ causes the lowest rate of impacts on the Earth.  There is a general trend for the total number of asteroid impacts on Earth for super-Earths located interior to Earth's orbit. As the separation of the super-Earth (from the Sun) increases, the asteroid impact flux on Earth also increases.

\begin{figure*}  \centering
\includegraphics[width=8.7cm]{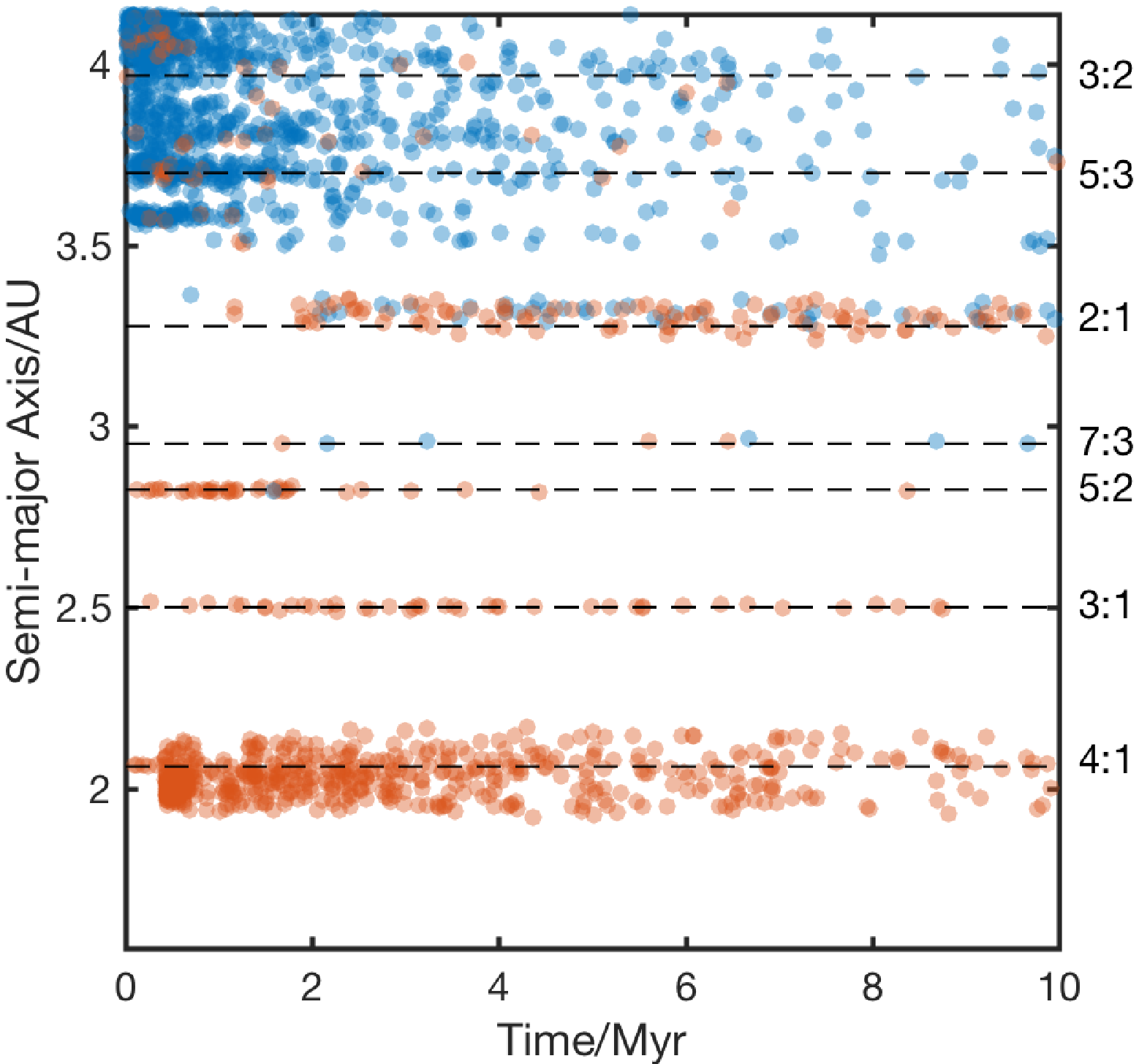}
\includegraphics[width=8.7cm]{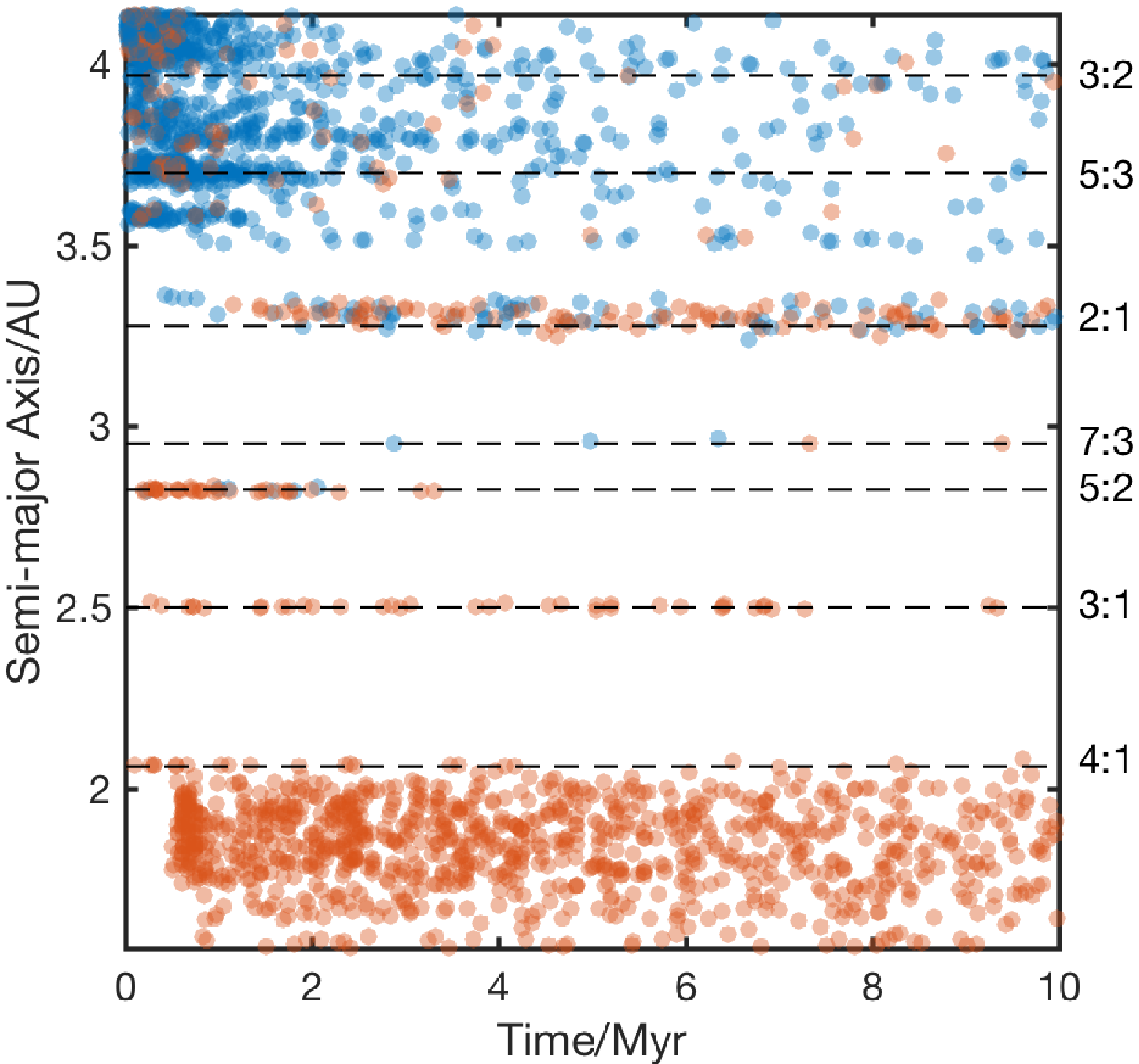}
\includegraphics[width=8.7cm]{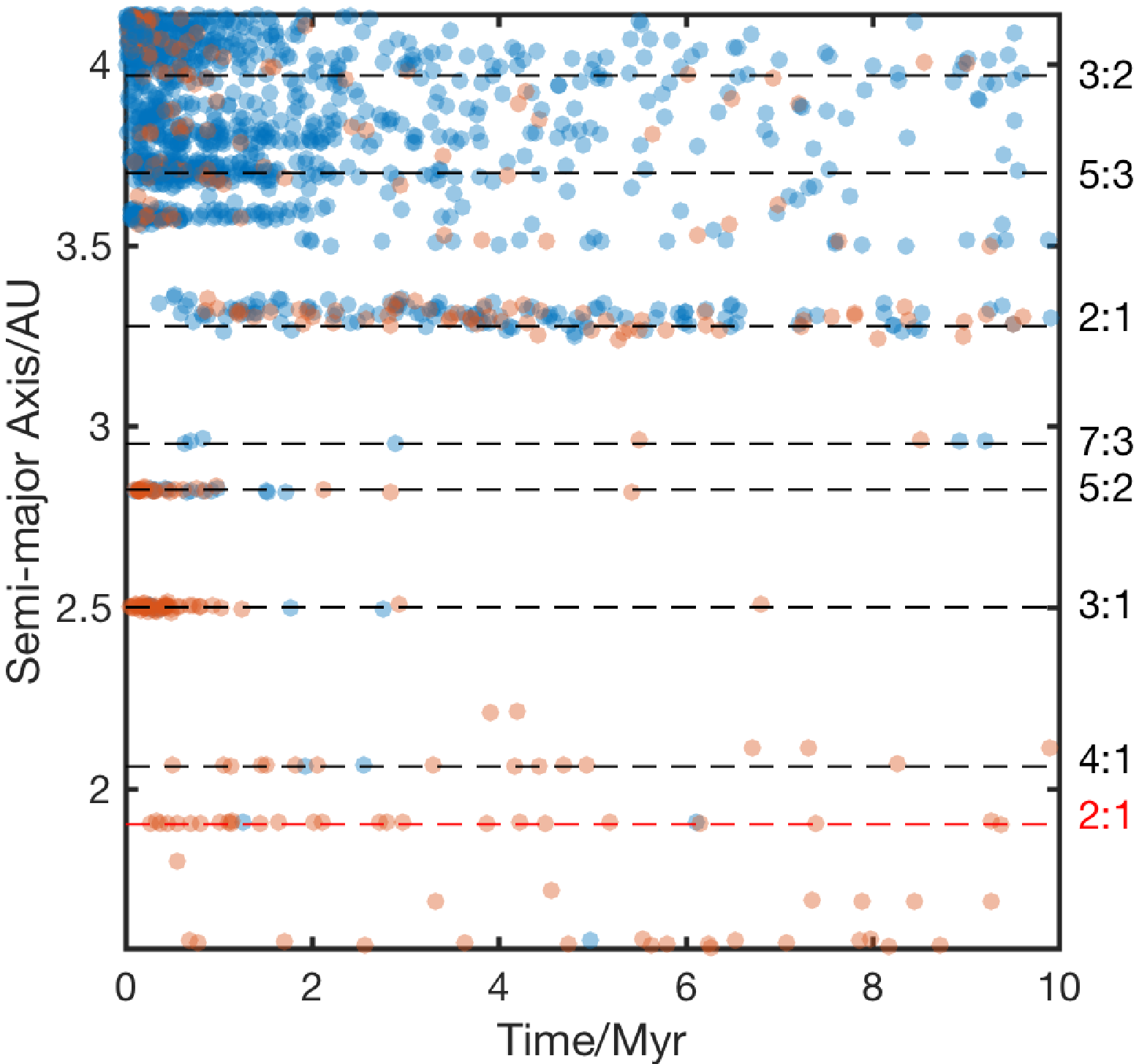}
\includegraphics[width=8.7cm]{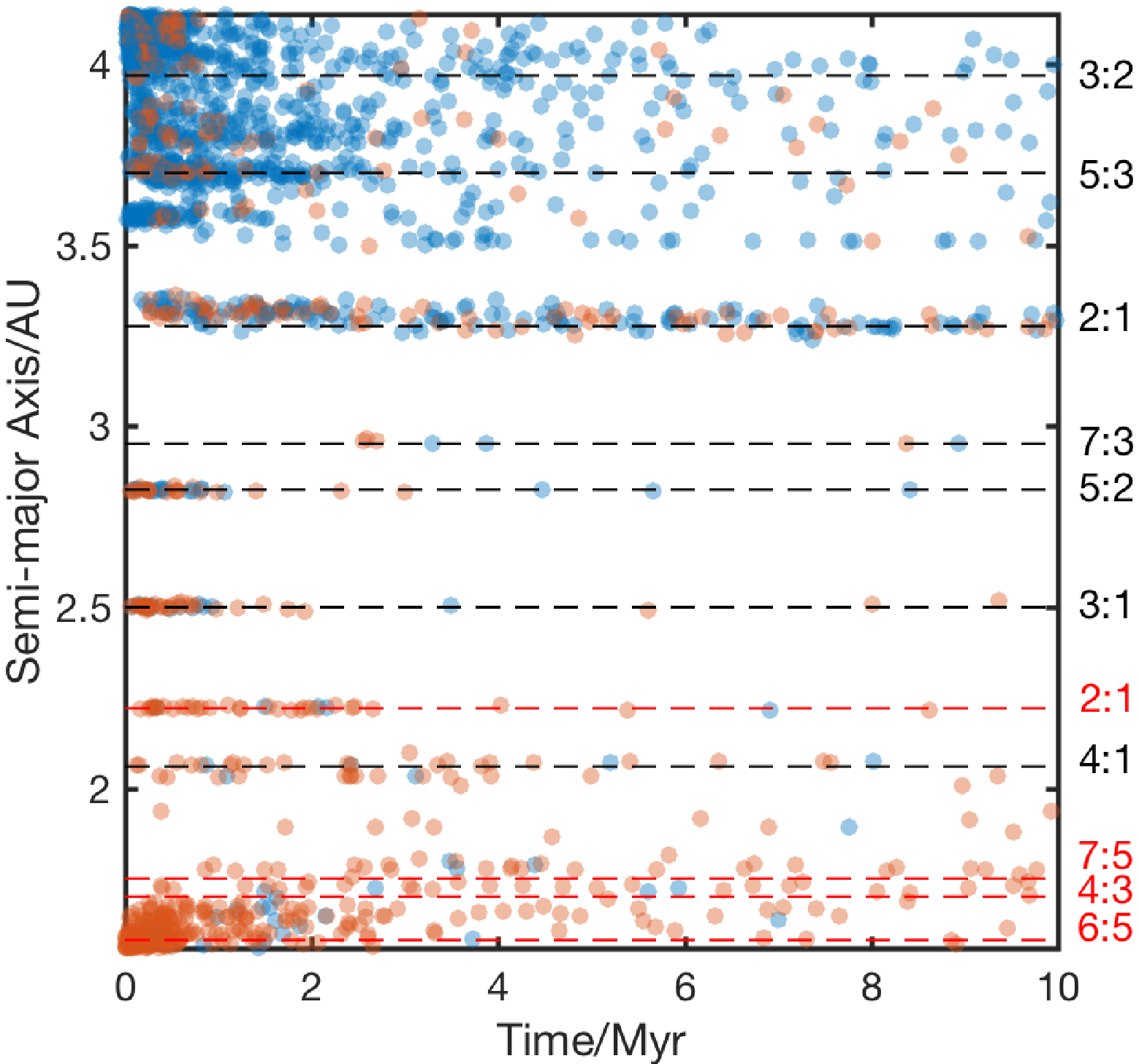}
\caption{The original semi-major axis of each asteroid as a function of the time when the final outcome occurred. The possible outcomes for each asteroid include Earth-impact (red) and other (blue). "Other" refers to ejection, super--Earth impact, Juptier--impact, Saturn--impact, or colliding with central object. The inner and outer boundary of the asteroid distribution are located at $a_{\rm min} = 1.558 \, \rm AU$ and $a_{\rm max} = 4.138 \, \rm AU$. \textit{Top-left Panel:} Asteroid outcomes for a system with no super--Earth. \textit{Top-Right Panel:} $ \rm 10\,M_{\oplus}$ super--Earth located at a semi-major axis of $0.8\, \rm AU$. \textit{Bottom-Left Panel:} $\rm 10\,M_{\oplus}$ super--Earth located at a semi-major axis of $1.2\, \rm AU$. \textit{Bottom-Right Panel:} $\rm 10\,M_{\oplus}$ super--Earth located at a semi-major axis of $1.4\, \rm AU$. The mean-motion resonances with Jupiter are represented with the black-dotted line, with the name of each resonance listed to the right of their respected line. The mean-motion resonances with the super--Earth are represented by the red-dotted lines.}    
\label{outcome}
\end{figure*}

\subsection{Evolution of the asteroid belt}

Figure~\ref{belt} shows the evolution of the asteroid belt for simulations involving no super--Earth (top-left panel), a $\rm 10\, M_{\oplus}$ super--Earth at $a = 0.8\, \rm AU$ (top-right panel), a $\rm 10\, M_{\oplus}$ super--Earth  at $a = 1.2\, \rm AU$ (bottom-left panel), and  a $\rm 10\, M_{\oplus}$ super--Earth at $a = 1.4\, \rm AU$ (bottom-right panel). Each one of these systems also contains Earth, Jupiter, and Saturn. The distribution was calculated every million years for ten million years. As the evolution takes place, areas of the asteroid distribution proceed to be cleared out by gravitational perturbations. These perturbations are the result of mean--motion and secular resonances with Jupiter and Saturn. The most notable mean--motion resonances are the 3:1, 5:2, 7:3, and 2:1, located at $2.5\, \rm AU$, $2.8\, \rm AU$, $2.9\, \rm AU$, and $3.3\, \rm AU$, respectively. These resonances are located at the same location as our Kirkwood Gaps. Jupiter's chaotic region is located from $3.6\, \rm AU$ to the outer boundary of the asteroid distribution ($4.133\, \rm AU$). This region is produced by overlapping widths of the mean--motion resonances \citep{Murray1997, Murray1999}.  The $\nu_{6}$ resonance is located at $2.0\, \rm AU$ along with Jupiter's 4:1 mean--motion resonance. The system containing a $\rm 10\, M_{\oplus}$ super--Earth located at a semi--major axis of $0.8\, \rm AU$ appears to have a broader $\nu_6$  width than a system with no super--Earth. In contrast, if the super--Earth is located exterior to Earth's orbit, the $\nu_6$ resonance completely disappears. We discuss this further in sections $3.3$ and $3.5$.

\begin{figure}  \centering
\includegraphics[width=8.4cm]{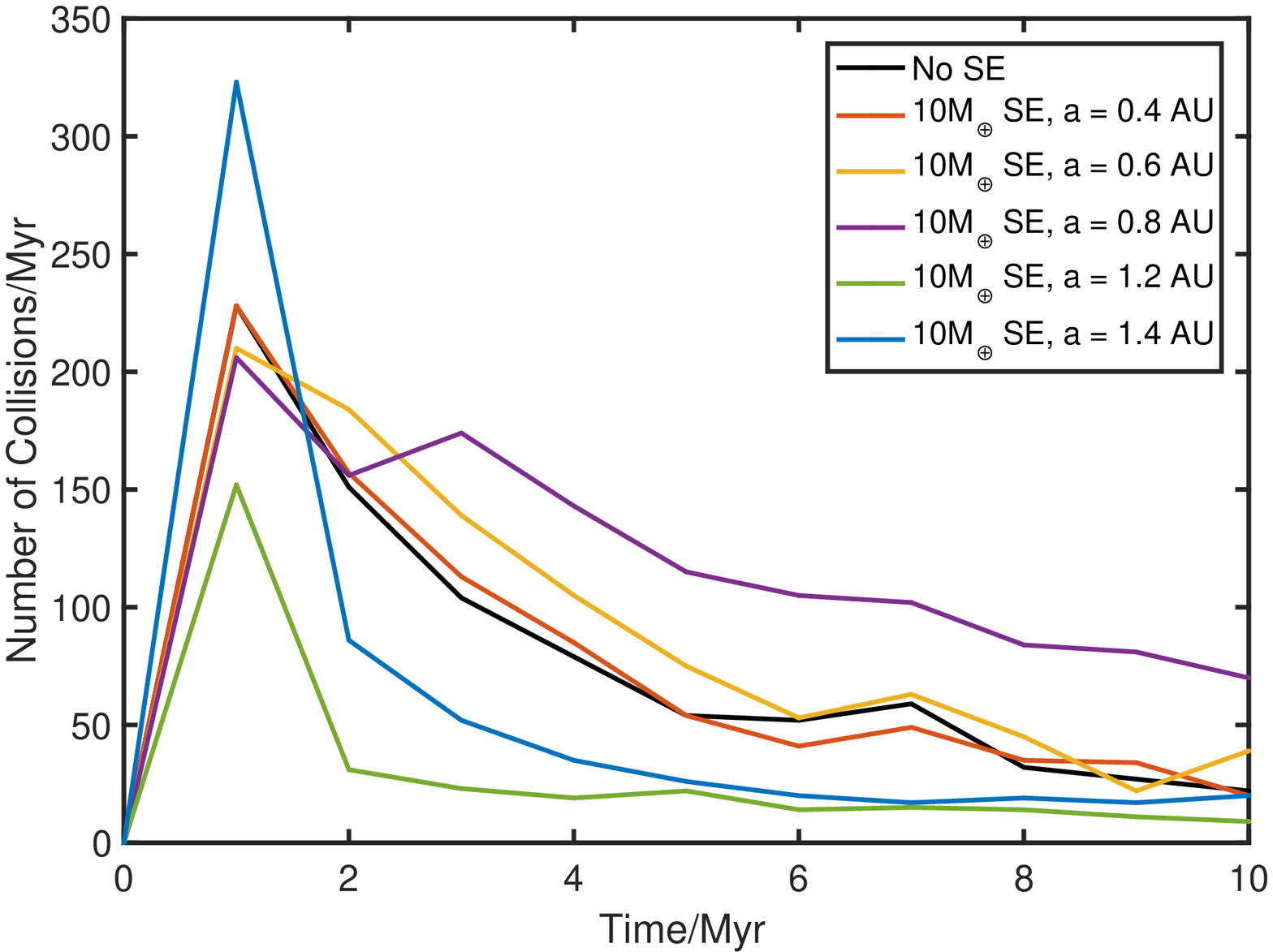}
\includegraphics[width=8.4cm]{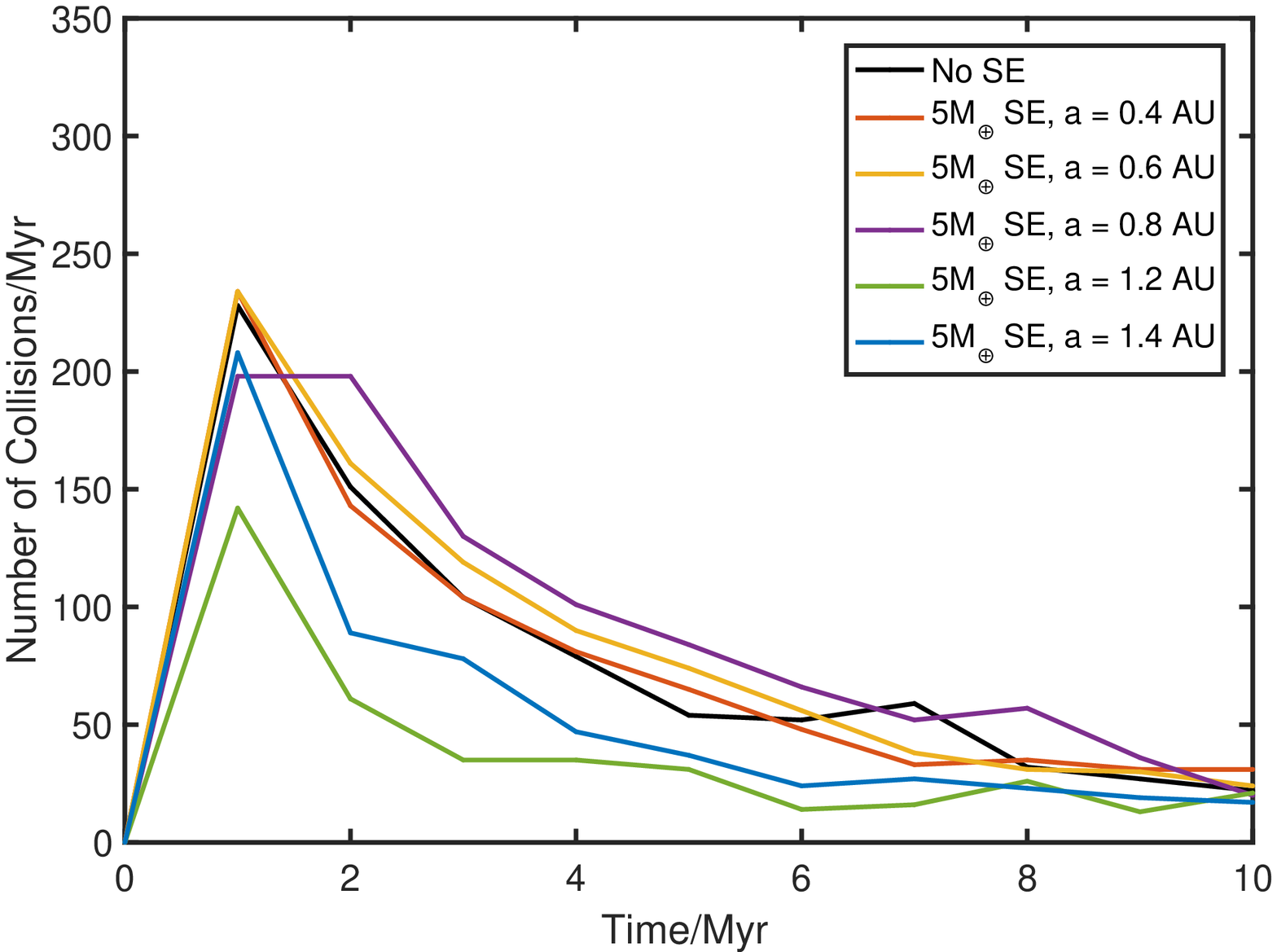}
\caption{Number of asteroid collisions towards Earth per million year for select simulations described in Table 1. \textit{Top Panel:} Collision rate for simulations involving a $\rm 10M_{\oplus}$ super--Earth. \textit{Bottom Panel:} Collision rate for simulations involving a $\rm 5M_{\oplus}$ super--Earth.}
\label{collision}
\end{figure}

\subsection{Resonances in the asteroid belt}

There are five potential outcomes for each asteroid during the various simulations. These are: ejection of the asteroid from the solar system, collision with the Earth, collision with another planet, collision with the Sun, or the asteroid remaining in the asteroid belt. Figure~\ref{outcome} shows the various asteroid outcomes for a system with no super--Earth (top--left panel), a  $10\,\rm M_\oplus$ super--Earth located at a semi--major axis $a = 0.8\,\rm AU$ (top--right panel),  a  $10\,\rm M_\oplus$ super--Earth located at $a = 1.2\,\rm AU$ (bottom--left panel), and a  $10\,\rm M_\oplus$ super--Earth located at $a = 1.4\,\rm AU$ (bottom--right panel). The initial semi-major axis of each asteroid is shown as a function of the time of its final outcome. The asteroids that were cleared out are originally located at the resonance locations, because  mean-motion and secular resonances operate over the initial semi-major axis of each asteroid, increasing the asteroid's eccentricity. On the right hand vertical axis we show the mean--motion resonance locations with Jupiter and the super-Earth. When a $10\,\rm M_\oplus$ super--Earth is located at $a = 0.8\,\rm AU$, there is a widening of the $\nu_{6}$ secular resonance, which  increases the number of asteroids perturbed onto Earth-colliding orbits. When the super--Earth is placed exterior to Earth's orbit, there is a substantial decrease in the number of Earth--colliding asteroids. This is due to the $\nu_6$ resonance being suppressed. When the super--Earth is located exterior to Earth's orbit, a 2:1 mean-motion resonance is created within the asteroid belt. This resonance causes additional asteroids to be cleared out. A chaotic zone due to the super--Earth,  is seen for the system with a $10M_{\oplus}$ super--Earth located at $1.4\, \rm AU$. This chaotic zone is produced from the overlapping libration widths of the super--Earth's 6:5, 4:3, and 7:5 mean-motion resonances. This chaotic zone causes a larger number of asteroids to be cleared out from the inner parts of the asteroid belt.

\begin{figure*}  \centering
\includegraphics[width=8.0cm]{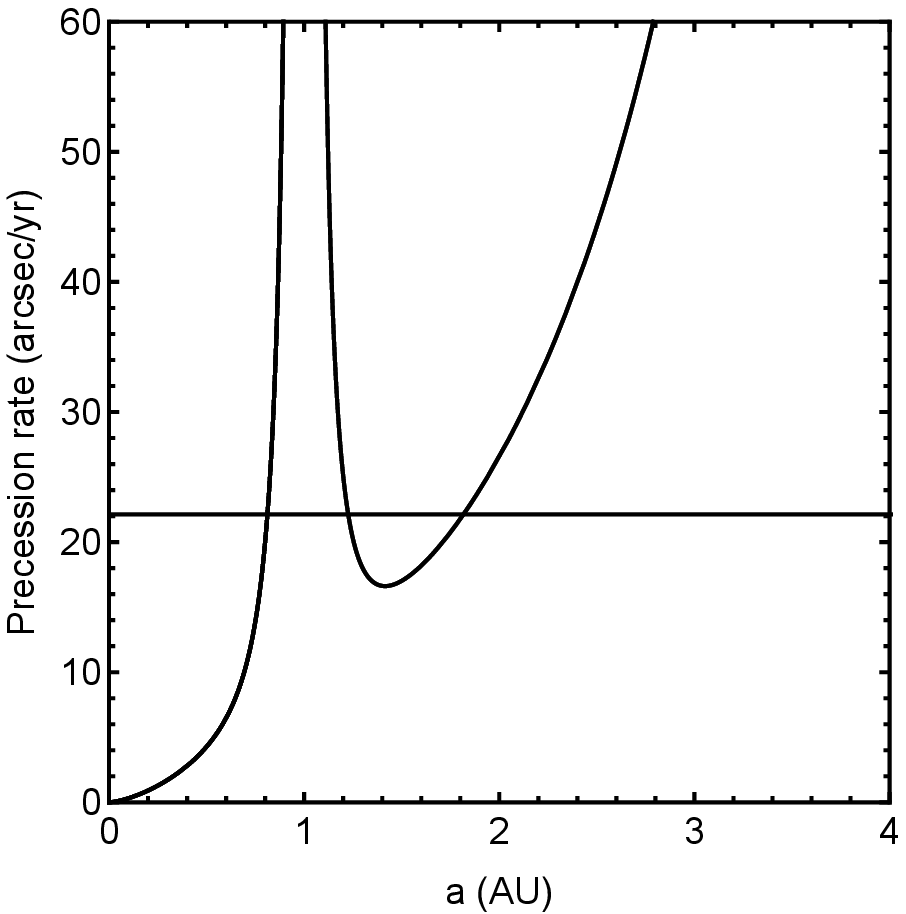}
\includegraphics[width=8.0cm]{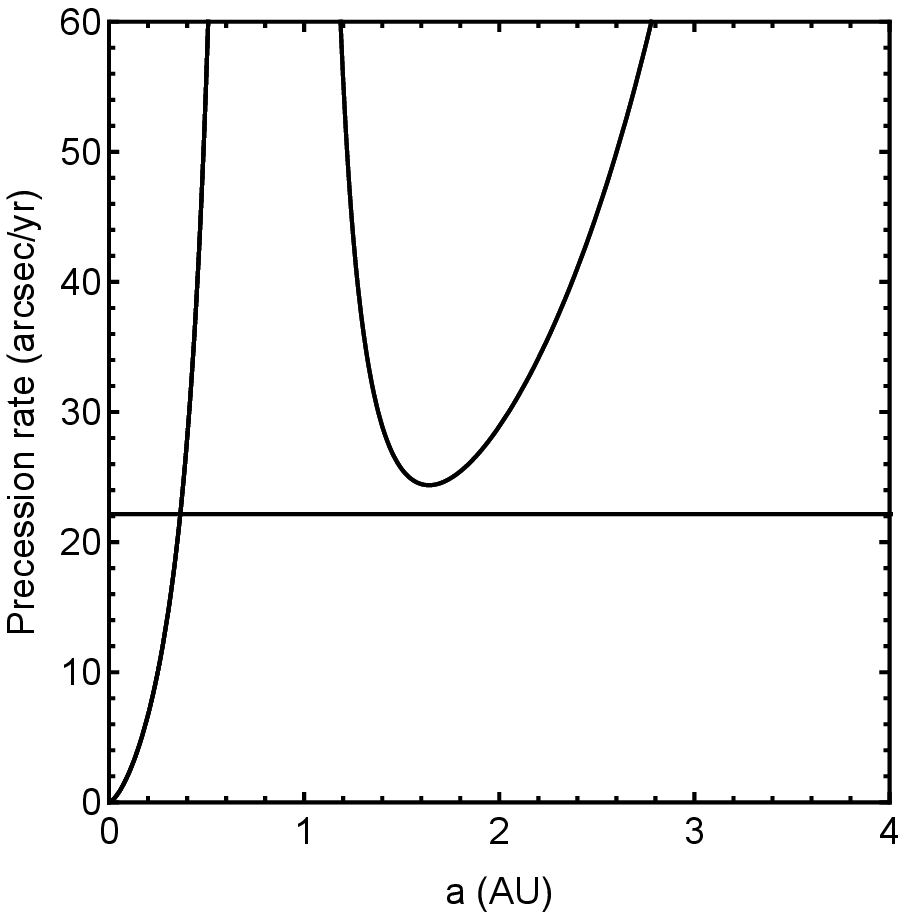}
\includegraphics[width=8.0cm]{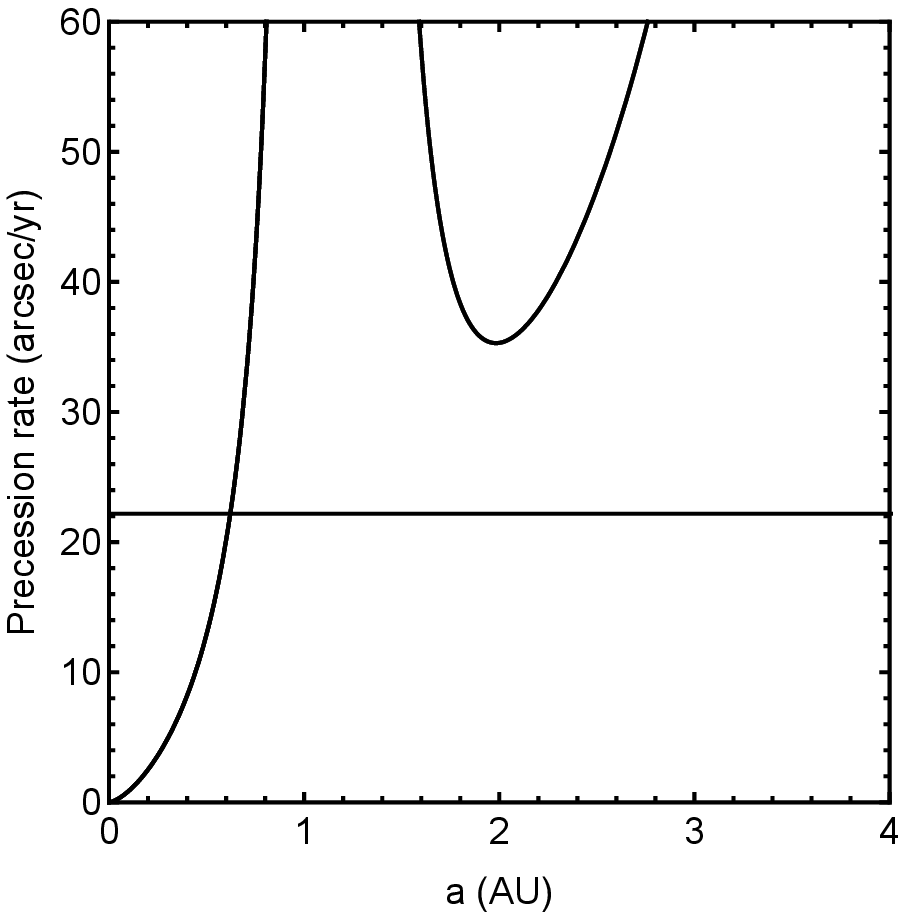}
\includegraphics[width=8.0cm]{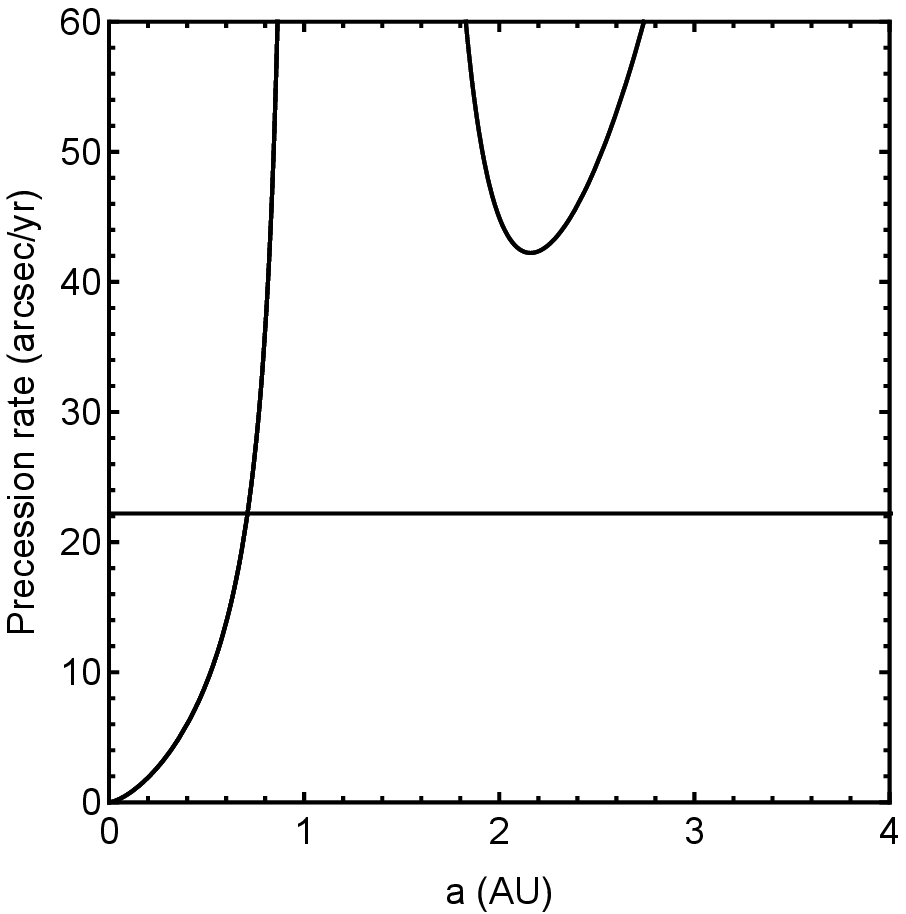}
\caption{The precession rate of a test particle as a function of semi-major axis in the inner part of the solar system. The solid horizontal line represents the $g_6$ eigenfrequency of Saturn. The intersection of the precession rate of the test particle with the eigenfrequency of Saturn denotes the location of a secular resonance. \textit{Top-Left Panel:} A system with no super--Earth. In this case, the intersection located at $2\, \rm AU$ is the location of the $\nu_6$ resonance within the asteroid belt. \textit{Top-Right Panel:} System with a $10\, \rm M_{\oplus}$ super--Earth located at a semi--major axis of $a = 0.8\, \rm AU$.  \textit{Bottom-Left Panel:} System with a $10\, \rm M_{\oplus}$ super--Earth located at a semi--major axis of $a = 1.2\, \rm AU$.  \textit{Bottom-Right Panel:} System with a $10\, \rm M_{\oplus}$ super--Earth located at a semi--major axis of $a = 1.4\, \rm AU$.} 
\label{p_rate}
\end{figure*}

\subsection{Collision rate on the Earth}

To determine the frequency of asteroid collisions with the Earth, the asteroid collision rate was calculated and plotted in Fig.~\ref{collision}.  In the top panel, we show the collision rate for systems with a  $10\,\rm M_\oplus$ super--Earth at various semi-major axis locations and in the bottom panel, the  collision rate for systems with a  $5\,\rm M_\oplus$ super--Earth for various semi-major axis values. The collision rate was calculated per million years, for ten million years. Initially, the collision rate is higher than compared to later times, due to the fact that there is a larger asteroid population at the beginning of the simulations than at the end. After one million years, the super--Earth located at $1.4\, \rm AU$ caused the greatest number of collisions, but the rate rapidly declined. This behavior is due to the chaotic region of the super--Earth, which quickly clears out asteroids in locations of overlapping resonances, perturbing them onto Earth--crossing orbits. This chaotic region can be clearly seen in the bottom--right panel in Fig.~\ref{outcome}. The system with a $10\,\rm M_\oplus$ super--Earth at $0.8$AU has the highest asteroid impact rate for the duration of the simulation. When comparing with the simulations that involve  $5\,\rm M_\oplus$ super--Earths, the collision rate is also the highest for a super--Earth located at $a = 0.8\,\rm AU$.

\subsection{The $\nu_6$ resonance}

As shown in Fig.~5, the majority of the asteroid collisions with the Earth are from asteroids that originate from the location of the $\nu_6$ resonance in the asteroid belt \citep[see also][]{Morbidelli1994,Gladman1997,Bottke2000,Minton2011}.  Thus, this secular resonance may have played a significant role in making the Earth habitable. In our simulation without a super--Earth, the total number of asteroids that collided with the Earth from secular resonances was about two and a half times that from mean--motion resonances. 

The $\nu_6$ secular resonance is due to both Saturn and Jupiter. Jupiter increases the precession frequency of the asteroids so that they fall into a resonance with the apsidal precession rate of Saturn.  In Fig.~\ref{p_rate} we show the precession rate of a test particle as a function of orbital separation. The solid horizontal line denotes the $g_6$ eigenfrequency of Saturn which is found by calculating the eigenvalues of the matrix $A_{ij}$ associated with a generalized form of the secular perturbation theory \cite[See, e.g., Section 7.7 in][]{MurrayBook2000}. The top-left and -right  correspond to the systems which include no super--Earth and a super--Earth located interior to Earth, respectively, and the bottom-left and -right both correspond to systems with a super-Earth located exterior to Earth. The values of Saturn's eigenfrequency for each of these systems are found to be $22.13^{\prime \prime} \rm yr^{-1}$, $22.15^{\prime \prime} \rm yr^{-1}$, $22.18^{\prime \prime} \rm yr^{-1}$, and $22.20^{\prime \prime} \rm yr^{-1}$, respectively. The inclusion of a super--Earth has little influence on the value of the $g_6$ eigenfrequency. We also find that the inclusion of Saturn does not noticeably affect the precession rate of the asteroids--that rate is dominated by Jupiter. The top-left panel of Fig.~\ref{p_rate} resembles our solar system with the inclusion of Earth, Jupiter, Saturn, and the asteroid belt. The intersection of the particle's precession rate with Saturn's eigenfrequency represents the location of the $\nu_6$ resonance, which is located at $\sim 2 \, \rm AU$.  

The inclusion of a super--Earth, which has a mass that is significantly smaller than that of the giant planets Jupiter and Saturn, can lead to a large enough change in the asteroid precession rate to enhance or to remove the resonance with Saturn, depending on the location {\textbf and mass} of the super--Earth. The enhancement of the $\nu_6$ resonance is observed in the top-right panel of Fig.~\ref{p_rate}, where the super--Earth is located interior to Earth's orbit at a semi-major axis $a = 0.80 \, \rm AU$. The precession rate of the test particle is close to the eigenfrequency of Saturn for semi-major axis values from $\sim 1.5 \, \rm AU$ to $\sim 2 \, \rm AU$, which causes an enhancement of the $\nu_6$ resonance.  

Re-examining the top-left panel in Fig.~\ref{p_rate}, there are two $\nu_6$ resonance locations, located at about $1.3\, \rm AU$ and $2.0\, \rm AU$, respectively. When a super--Earth is included within this model at an orbital separation of $0.8\, \rm AU$, these two resonance locations become closer together as the mass of the super--Earth increases. We follow \cite{malhotra2012} by constructing an analytical toy model of the $\nu_6$ resonance width by calculating the forced maximum eccentricity of a test particle near the $\nu_6$ resonance with and without a super--Earth. When the mass of the super--Earth reaches $10\, \rm M_{\oplus}$, the widths of the two $\nu_6$ resonances overlap and the resonance  operates over a larger range of semi-major axis values.

The removal of the $\nu_6$ resonance occurs when the super--Earth is located exterior to Earth's orbit. This can be seen in the bottom-left panel (super--Earth located at $a = 1.2 \, \rm AU$) and the bottom-right panel (super--Earth located at $a = 1.5 \, \rm AU$) of Fig.~\ref{p_rate}, where the precession rate of the test particle does not intersect with the eigenfrequency of Saturn, leading to the disappearance of the $\nu_6$ resonance. This analysis of the precession rate of a test particle agrees with the results described in the previous Section. Re-examining Fig. ~\ref{outcome}, we can see that when the super--Earth is located interior to Earth's orbit we have a widening of the width ($\sim 1.5 \, \rm AU$ to $\sim 2.0 \, \rm AU$) of the $\nu_6$ resonance, whereas if the super--Earth is located exterior to Earth's orbit, the resonance is removed. Thus, the enhancement or removal of the $\nu_6$ resonance is predicated on the location {\textbf and mass} of the super--Earth. 

\begin{table*}
	\centering
	\caption{Number of asteroidal outcomes with varying parameters of Saturn. Each simulation includes  Earth, Jupiter (J), and Saturn (S). Note that since the size of the Earth has been inflated, the number of Earth impacts cannot be compared to the other outcomes. We only make comparison between the number of Earth impacts between different simulations.
	}
	\label{tab:example_table}
	\begin{tabular}{cccccccc} 
		\hline
		Simulation Name & Saturn Mass & Saturn Semi-major Axis & Earth Impacts &J/S Impacts & Star Collision & Ejected & Remaining\\
        & $\rm M_S$ & AU & & & & \\
\hline
run22 & $1.0$ & $8.0$ & 514 & 113 & 2 & 2705 & 6666 \\
run23 & $1.0$ & $9.0$ & 1012 & 37 & 0 & 1206 & 7745 \\
run24 & $1.0$ & $9.537$ & 808 & 55 & 2 & 1112 & 8023 \\
run25 & $1.0$ & $10.0$ & 957 & 48 & 4 & 1034 & 7957 \\
run26 & $1.0$ & $11.0$ & 248 & 50 & 1 & 966 & 8735 \\
run27 & $1.0$ & $12.0$ & 210 & 54 & 1 & 928 & 8807 \\
\hline
run28 & $0.1$ & $9.537$ & 172 & 83 & 1 & 874 & 8870 \\
run29 & $0.5$ & $9.537$ & 447 & 55 & 2 & 1005 & 8491 \\
run30 & $1.0$ & $9.537$ & 808 & 55 & 2 & 1112 & 8023 \\
run31 & $1.5$ & $9.537$ & 1163 & 49 & 3 & 1238 & 7547 \\
\hline
	\end{tabular}
    \label{table2}
\end{table*}

\begin{figure*}  \centering
\includegraphics[width=8.4cm]{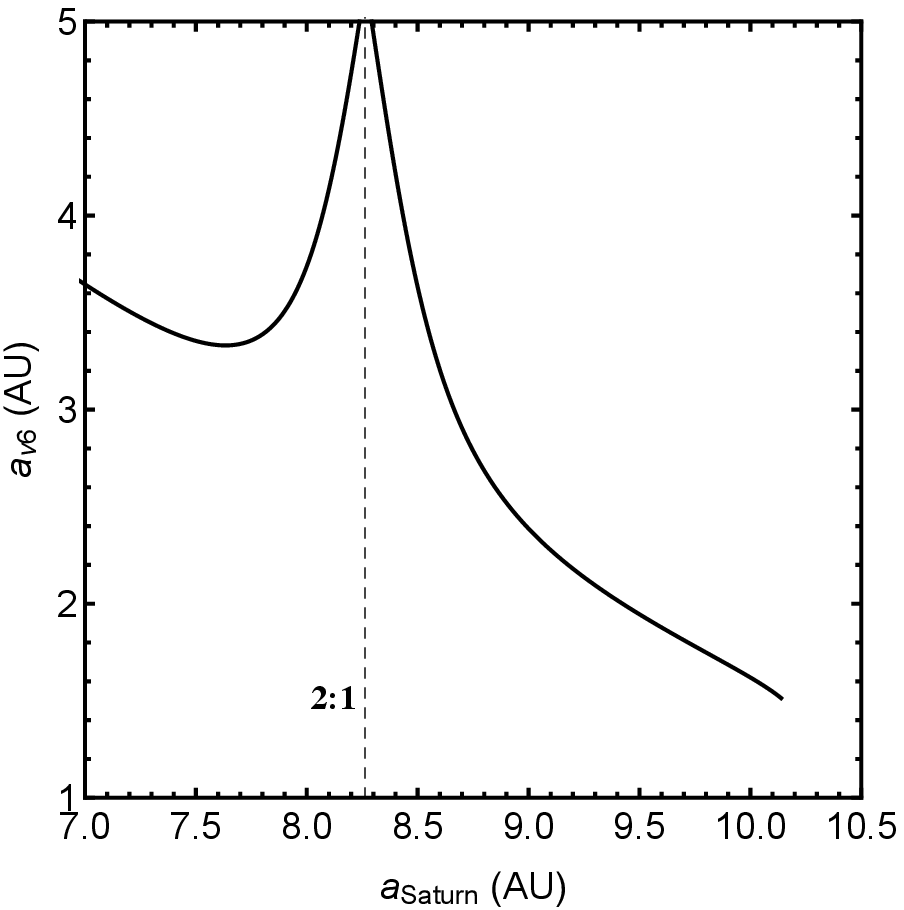}
\includegraphics[width=8.4cm]{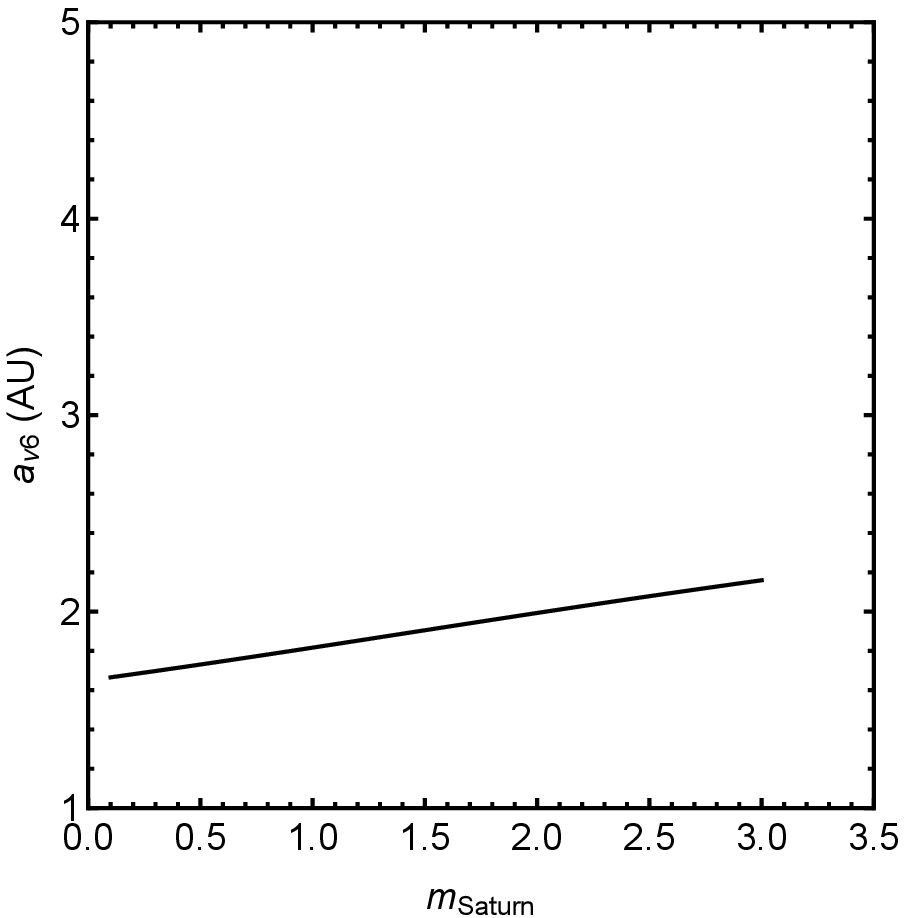}
\caption{\textit{Left Panel:} Location of the $\nu_6$ secular resonance with respect to the semi-major axis of Saturn. We have included a correction due to the 2:1 mean-motion resonance between Jupiter and Saturn \citep{Malhotra1989,Minton2011}. The dotted line shows the location of this 2:1 mean-motion resonance. \textit{Right Panel:} Location of the $\nu_6$ resonance as a function of the mass of Saturn.}   
\label{nu6}
\end{figure*}

\section{The architecture of the outer solar system}

\begin{figure*}  \centering
\includegraphics[width=8.7cm]{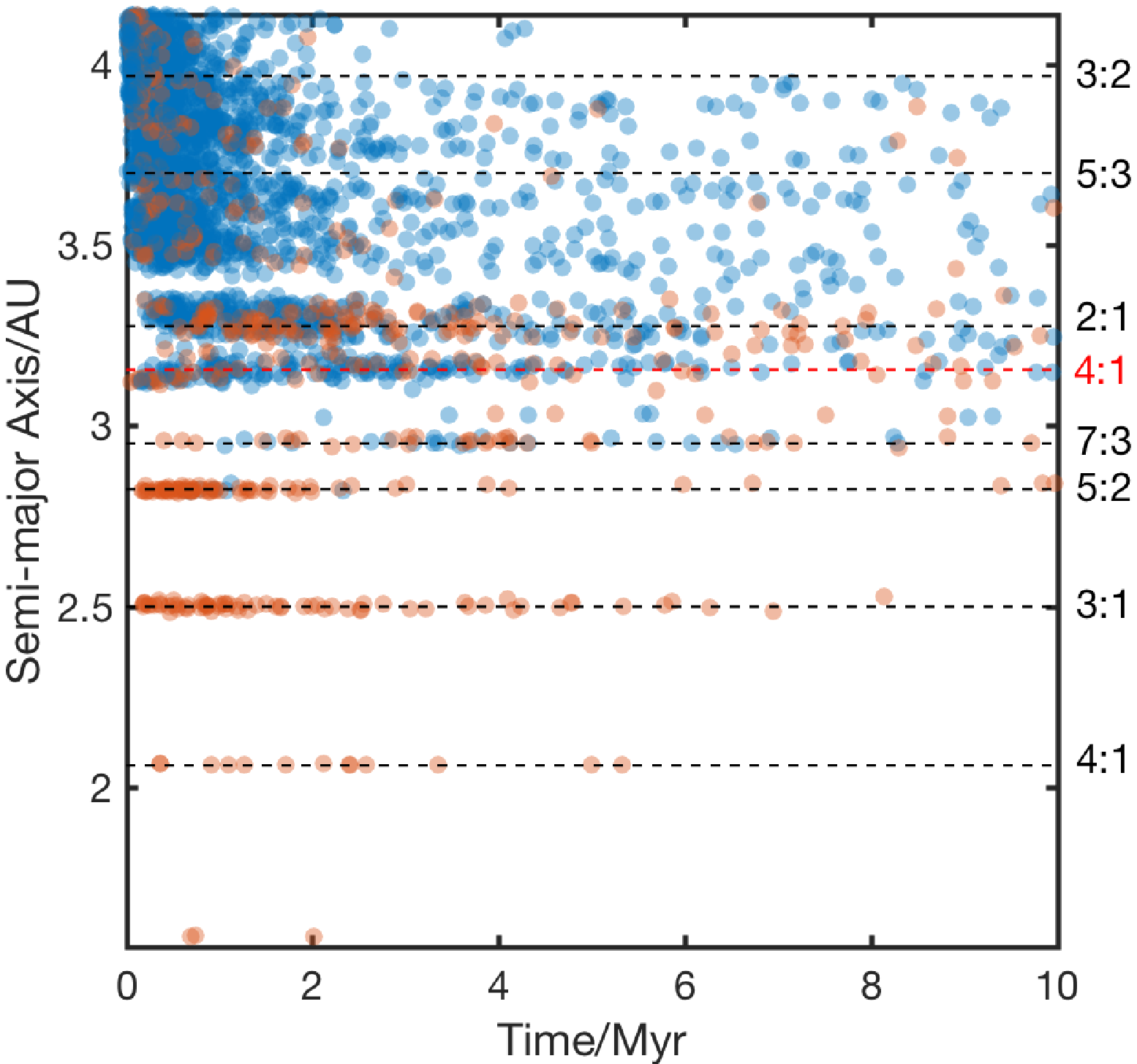}
\includegraphics[width=8.7cm]{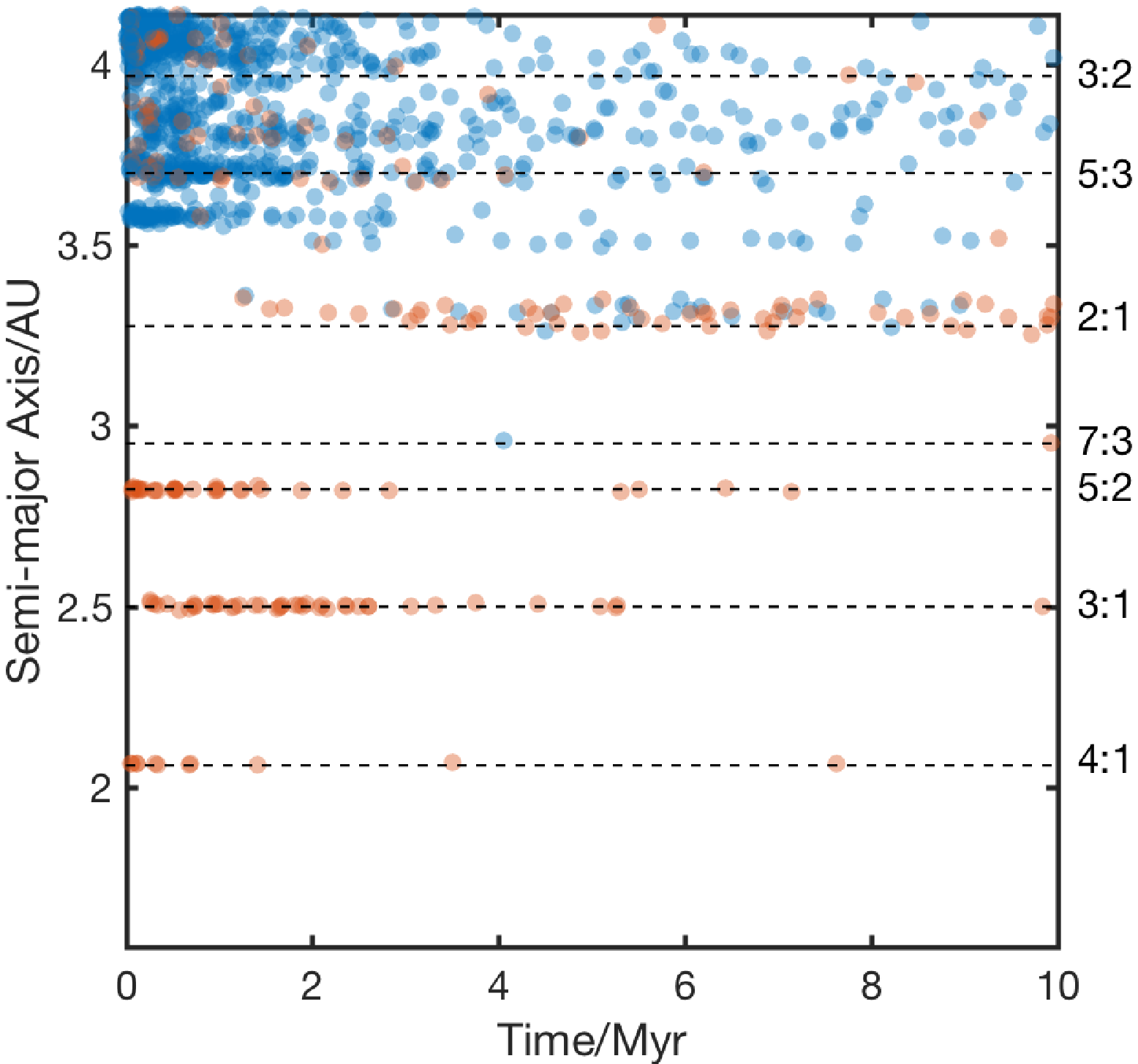}
\caption{The original semi-major axis of each asteroid as a function of the time when the final outcome occurred. The possible outcomes for each asteroid include Earth-impact (red) and other (blue). "Other" refers to ejection, Juptier--impact, Saturn--impact, or colliding with central object. The inner and outer boundary of the asteroid distribution are located at $a_{min} = 1.558 \, \rm AU$ and $a_{max} = 4.138 \, \rm AU$. \textit{Left Panel:} Asteroid outcomes for a system with Saturn located at a semi-major axis $a = 8 \, \rm AU$. \textit{Right Panel:} Saturn located at a semi--major axis $a = 12 \, \rm AU$. The mean--motion resonances with Jupiter are represented with the black--dotted line and the red-dotted line represents a mean-motion resonance between the asteroids and Saturn. Each mean-motion resonance is listed to the right of their respected line.}   
\label{saturn}
\end{figure*}

We next considered the properties of the outer giant planets that produce a $\nu_6$ resonance in the location of the asteroid belt.  We keep the mass and orbital separation of Jupiter fixed. The location of the $\nu_6$ resonance as a function of Saturn's semi-major axis and mass is shown in Fig.~\ref{nu6}.  In the left panel, the location of the $\nu_6$ resonance as a function of Saturn's semi-major axis was found by calculating the resulting eigenfrequency of Saturn and then finding the location of the intersection with the precession rate of a test particle. The precession frequency of the test particle was calculated by including the planets  Earth, Jupiter, and Saturn. However,  Saturn's eigenfrequency was calculated by including only Jupiter since the Earth has a negligible influence in the calculation. We included a correction due to the near 2:1 mean-motion resonance between Jupiter and Saturn \citep{Malhotra1989,Minton2011}. The location of the resonance drastically moves outwards as Saturn becomes closer to a 2:1 resonance with Jupiter at about $8.3\,\rm AU$. As Saturn moves outwards, the resonance location moves inwards. 
The right panel  of Fig.~\ref{nu6} shows the location of the $\nu_6$ resonance as a function of Saturn's mass, while Saturn is located at its current semi-major axis of $9.5\,\rm AU$. We direct the readers to a more accurate calculation of \cite{Minton2011}(see their figure 2) for the value of $g_6$ and the $\nu_6$ location as a function of Saturn's semi-major axis. If we increase Saturn's mass, then the location of the resonance moves outwards slightly. For a mass that is one tenth of its current value, the resonance moves inwards only slightly. The resonance location is rather more sensitive to the orbital separation than to the mass of Saturn. 

We have run additional numerical simulations of the asteroid belt to consider the effects of varying the properties of the outer giant planets. We varied Saturn's semi-major axis and mass to verify the trends found in Fig.~\ref{nu6}. The results of these simulations are presented in Table~\ref{table2}. Varying Saturn's semi--major axis (runs 22-27 from Table~\ref{table2}) follows the trend calculated in Fig.~\ref{nu6} (left panel). As Saturn's semi-major axis increases, the location of the $\nu_6$ resonance moves outside the asteroid belt boundaries. On the other hand, when the semi--major axis decreases, the $\nu_6$ resonance moves towards the middle/outer regions of the asteroid belt. This change in the location of the resonance can be seen in Fig.~\ref{saturn}, where the left panel represents the various asteroid outcomes when Saturn is located at a semi--major axis of $a = 8.0\, \rm AU$ and the right panel shows the outcomes for when Saturn is located at $a = 12.0\, \rm AU$. Again, the time of the particular outcome is plotted with respect to the initial semi-major axis of each asteroid. 

For the case where Saturn is located at $a = 8.0 \, \rm AU$, the location of the $\nu_6$ resonance is shifted outward to approximately $3.5 \, \rm AU$, which is what is expected according to the left panel of Fig.~\ref{nu6}. This movement of the resonance causes a decrease in the number of collisions with the Earth, but causes a significant increase in the number of asteroid ejections. For the situation where Saturn's semi-major axis is taken to be $12.0\, \rm AU$, there is no $\nu_6$ secular resonance. The disappearance of the resonance is caused by there being no intersection of the particles precession rate with Saturn's eigenfrequency. Overall we find that varying the orbital separation of Saturn in both directions has a significant effect on the location of the $\nu_6$ secular resonance and that it generally leads to a decrease in the number of asteroid collisions. However, the orbital location of Saturn may not be accidental, since it is close to being in the 5:2 resonance with Jupiter. \cite{Fernandez1984,Fernandez1996} suggested the jovian planets proximity to the current near resonant structure could be a consequence of the differential expansion of their orbits during late stages of planetary formation \cite[e.g.][]{Michtchenko2001}.

When the mass of Saturn is changed (runs 28-31 from Table~\ref{table2}), the calculated trend in Fig.~\ref{nu6} (right panel) breaks down for masses greater than $1.5 \, \rm M_{Saturn}$ (results not shown in Table~\ref{table2}). For the situations where Saturn's mass is greater than $1.5 \, \rm M_{Saturn}$, the Earth's orbit becomes highly eccentric ($e \approx 0.5$) which causes a higher number of asteroid impacts. When Saturn has a mass of $0.1 \,\rm M_{Saturn}$ the number of asteroid collisions with Earth is substantially lower, since the $\nu_6$ resonance is located outside the asteroid distribution. The location of the $\nu_6$ resonance is more sensitive to changes in Saturn's orbital separation than to its mass.

\section{Conclusions}

We found that the $\nu_6$ resonance may play an important role in producing asteroid collisions with terrestrial planets in the inner parts of a planetary system \citep[see also][]{Morbidelli1994,Gladman1997,Bottke2000,Ito2006}. Using N--body simulations and analytical models we have modeled planetary systems with terrestrial planets, an asteroid belt and two giant planets, in this order in terms of their separation from the central star. We have explored how the planetary system architecture affects the location and width of the $\nu_6$ resonance and how this in turn affects the number of asteroid collisions with the Earth.  Since super--Earths are common in the inner regions of exoplanetary systems we first considered their influence.  A super--Earth with a mass of around $10\,\rm M_\oplus$ at an orbital radius greater than about $0.7\,\rm AU$ may significantly affect the number and rate of asteroid collisions on the Earth. A super-Earth interior to the Earth's orbit increases the number of asteroid collisions with the Earth, while a super--Earth between the Earth and the asteroid belt decreases the number of asteroid collisions. Furthermore, we find that increasing the mass of Saturn increases the number of asteroid collisions. Changing the location of Saturn generally leads to a significant decrease in the number of asteroid collisions. The slight differences between our numerical calculations and analytical models were due to neglecting higher-order terms within our analytical models. 

\cite{schlichting2012} proposed that additional collisions from a residual planetesimal population are needed to damp the high eccentricities and inclinations of terrestrial planets during the late stages of planet formation. This damping mechanism would allow terrestrial planets to evolve onto circular and coplanar orbits.  Geochemical evidence suggests that the Earth accreted roughly $0.3\%$--$0.7\%$ of its total mass in the form of chondritic material during the late stages of planet formation \citep{Walker2009}. If these series of events are indeed true, this implies that asteroid collisions may have an important effect on determining the habitability of planets located in exoplanetary systems, which would make the results in the present work potentially very significant. In particular, and to end on a speculative note,  \cite{MartinLivio2016} proposed a model in which super--Earths did form in the inner parts of the solar system, but then migrated into  the Sun. Since we have shown that the rate of asteroid impacts on Earth is affected by the presence (or not) of an inner super--Earth, this chain of events, if it had indeed happened, surely had an effect on both the rate and pattern of asteroid impacts. It is tempting to speculate, therefore, that the resulting pattern and rate may have made the Earth more conducive to the emergence and evolution of life.

\section*{Acknowledgments}
All the simulations ran at the UNLV National Supercomputing Institute on the Cherry Creek cluster. This research made use of the Exoplanet Orbit Database and the Exoplanet Data Explorer at exoplanets.org.




\bibliographystyle{mnras}
\bibliography{main} 

\bsp	
\label{lastpage}
\end{document}